\documentclass[twocolumn,showpacs,floatfix,longbibliography]{revtex4-1}
\usepackage{graphicx}
\usepackage{bm} 
\usepackage{amssymb} 
\usepackage{amsmath} 
\usepackage{braket} 
\usepackage{natbib} 
\usepackage{hyperref}

\begin{document}

\title{Skyrmion-induced bound states in a superconductor}

\author{Sergey S. Pershoguba$^{1,3}$}
\author{Sho Nakosai$^{2,3}$}
\author{Alexander V. Balatsky$^{1,3}$}
\affiliation{$^1$Institute for Materials Science, Los Alamos National Laboratory, Los Alamos, New Mexico 87545, USA}
\affiliation{$^2$Condensed Matter Theory Laboratory, RIKEN, Wako, Saitama, 351-0198, Japan}
\affiliation{$^3$Nordita, Center for Quantum Materials, KTH Royal Institute of Technology, and Stockholm University, Roslagstullsbacken 23, S-106 91 Stockholm, Sweden}

\date{\today}

\begin{abstract}
We consider a superconductor proximity coupled to a two-dimensional ferromagnetic film with a skyrmion texture. Using the T-matrix calculations and numerical modeling we calculate the spin-polarized local density of states in the superconductor in the vicinity of the skyrmion. We predict the skyrmion bound states that are induced in the superconductor, similar to the well-known Yu-Shiba-Rusinov (YSR) states. The bound state wavefunctions have spatial power-law decay. It is suggested that superconductivity could facilitate an effective long-range interaction between skyrmions when bound state wavefunctions overlap.
\end{abstract}

\pacs{12.39.Dc, 74.45.+c}


\maketitle
\section{Introduction} \label{sec:intro}

Skyrmions, topological particle-like configurations of a continuous vector field, were originally proposed in the context of high-energy physics \cite{Skyrme}.  Nevertheless, it was suggested theoretically \cite{Bogdanov1989,Rossler2006} and recently confirmed experimentally \cite{Muhlbauer2009,Munzer2010,Yu2011,Heinze2011,Seki2012} that skyrmions exist in chiral ferromagnets in the presence of Dzyaloshinkii-Moriya interaction. Due to non-trivial topological properties, skyrmions manifest anomalous transport response to temperature gradients \cite{Jonietz2010} and electric field \cite{Neubauer2009,Zang2011,Saxena2013}. Recently, Hamburg group demonstrated a controllable writing and deleting of single skyrmions on the surface of PdFe bilayer \cite{Romming2013,Bergmann2014,Romming2015}.  Skyrmions hold a great promise in applications such as spintronics, memory devices, etc \cite{Fert2013,Nagaosa2013}. For example, interplay of a magnetic skyrmion and a topological insulator was recently considered in Ref.~\cite{Hurst2015}. Coupling of magnetic films with skyrmions to novel materials may produce new functionalities in hybrid devices not available in the constituent materials taken separately.

In parallel, there has been a significant interest in superconductor-ferromagnet (SC-FM) heterostructures aimed at engineering topological superconductors \cite{Alicea2012,Beenakker}. Discovery of the topological superconductivity would entail existence of the Majorana edge modes necessary for realizing topological quantum computing \cite{Nayak2008}. Motivated by the interest in skyrmions as well as SC-FM heterostructures  we connect the two fields in the current work.

\begin{figure} \centering
(a) \includegraphics[width=0.4\linewidth]{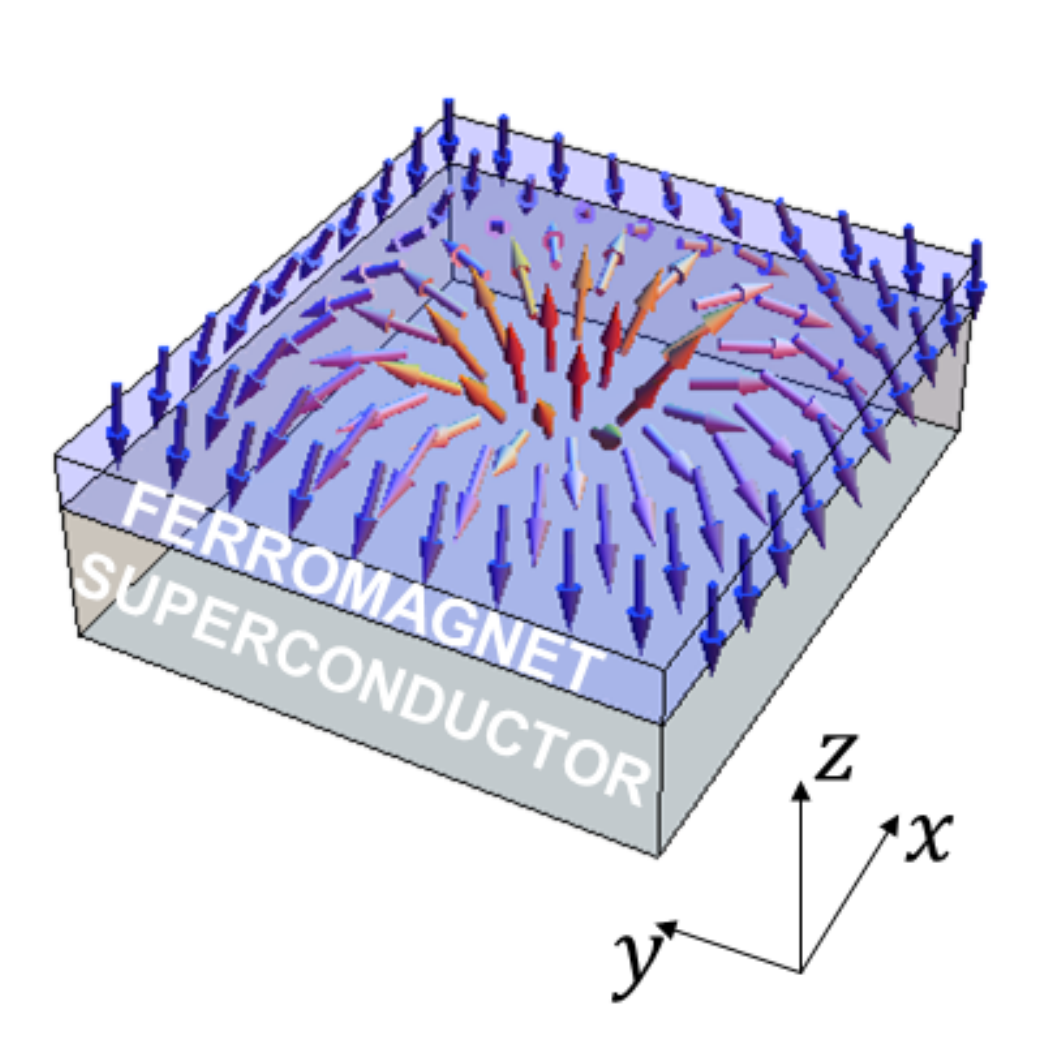}
(b) \includegraphics[width=0.4\linewidth]{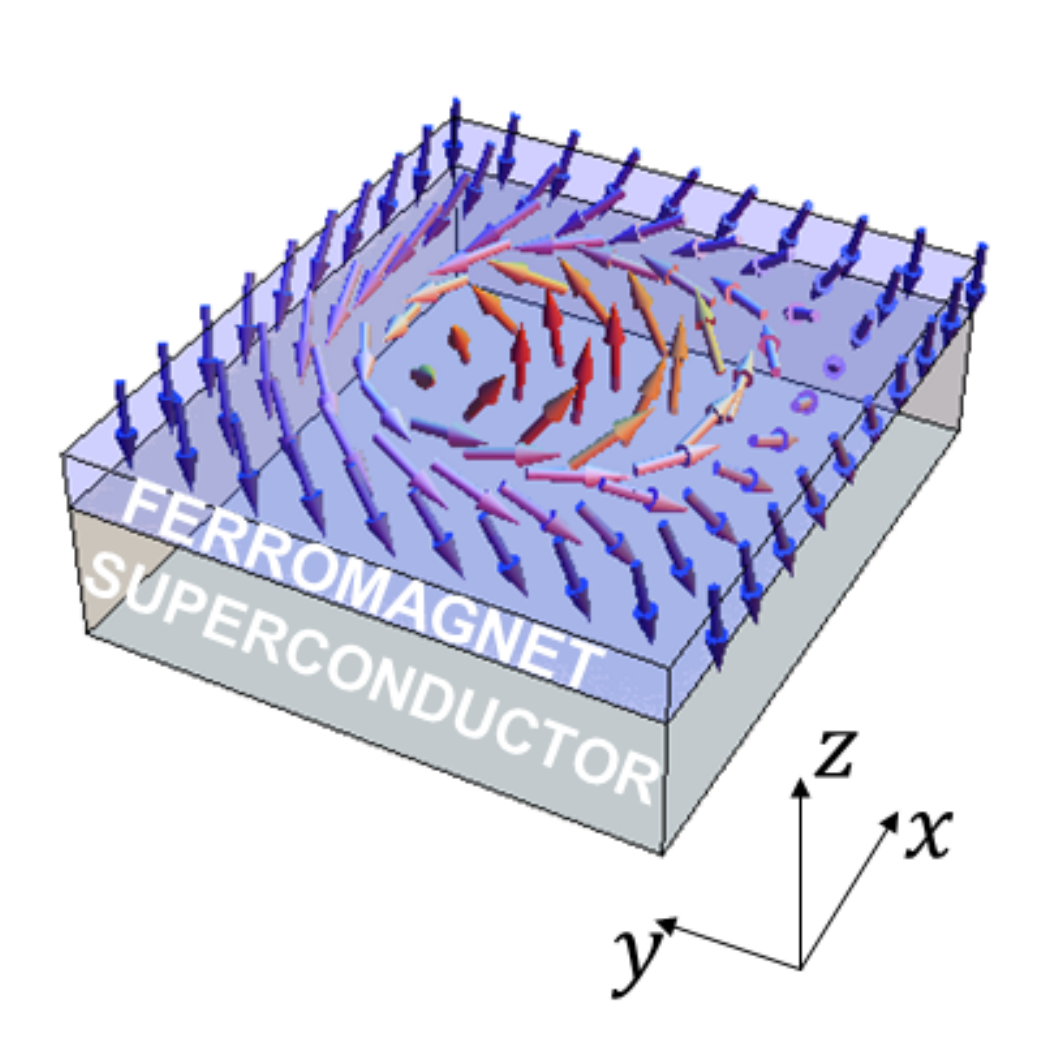} \\
(c) \includegraphics[width=0.4\linewidth]{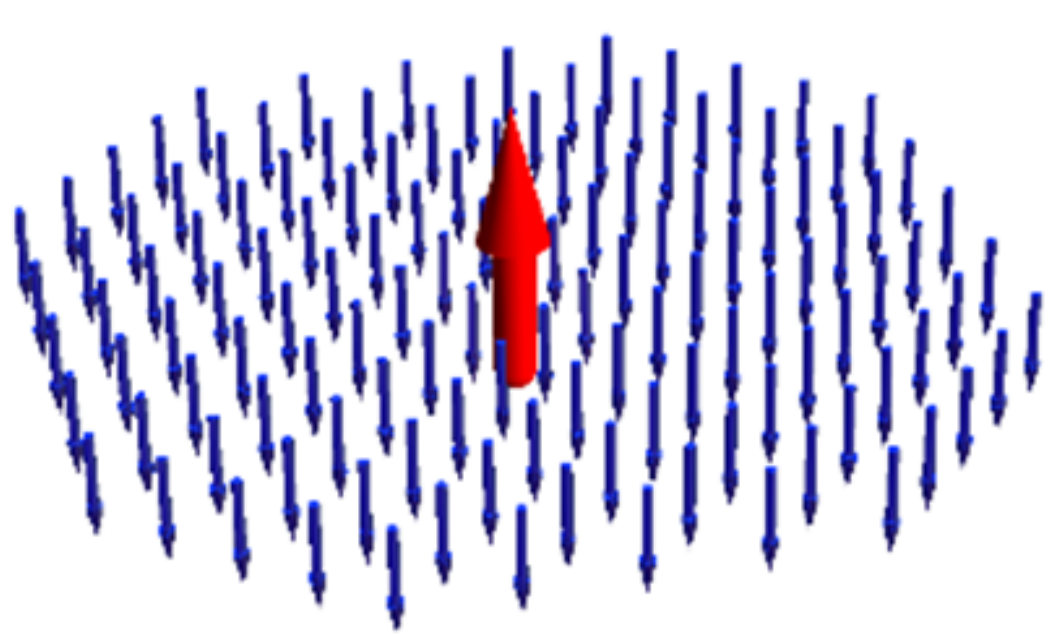}
\caption{(color online) (a,b) System under consideration: ferromagnetic (FM) film with a skyrmion proximity coupled to a superconductor (SC). (a) N\'eel-type skyrmion.  (b) Bloch-type skyrmion. (c) Sketch of an approximation of a skyrmion as a local magnetic moment floating in a ``ferromagnetic sea''.} \label{fig:skyrmion}
\end{figure}

Below, we consider a FM film with a skyrmion proximity coupled to SC as shown in Fig.~\ref{fig:skyrmion}. We search for the states in SC localized around a skyrmion in a series of approximations. First, consider a limit of a small skyrmion, i.e. $R\ll \xi_{\rm sc}$.  In this case, the approximation of the skyrmion field as a point magnetic moment is valid. Using this simplified model, we perform an analytical T-matrix calculation and find that skyrmion induces a bound state in the SC in a close analogy with the well-known Yu-Shiba-Rusinov states \cite{Yu,Shiba,Rusinov,Balatsky2006}. The bound state induces a resonance with a finite spectral width in a spin-polarized local density of states (SP LDOS). In contrast with the conventional YSR states, which are short-range, the skyrmion bound state is a long-range state with a power law decay. Therefore, in the presence of multiple skyrmions, the SC could mediate an effective long-range interaction between the skyrmions \cite{Yao2014} when the bound state wavefunctions overlap. Subsequently, we relax the requirement $R\ll \xi_{sc}$ and calculate the LDOS and wavefunctions for $R\sim \xi_{sc}$ numerically. We find that the bound state peak in the density of states is populated by the multiple quasilocalized states corresponding to different angular momenta.

We also note that a few earlier papers have considered skyrmions in the context of superconductivity to some extent. Reference~\cite{Garaud2011} studied the skyrmion-like solitons in the multiband superfluids and SCs. Paper~\cite{Nakosai2013} discussed a possibility of realizing a topological SC using a skyrmion lattice. The Josephson current through a magnetic skyrmion structure was considered in Ref.~\cite{Yokoyama2015}. None of the papers up-to-date have addressed the conceptually simplest case of interaction between a single skyrmion and SC. This is the subject of the present paper.

\section{Model: s-wave superconductor proximity coupled to a ferromagnetic film with a skyrmion} \label{sec:skyrmion}

Consider a FM film with the magnetization described by a three-dimensional vector $\bm S(\bm r) = (S_x,S_y,S_z)$ dependent on a two-dimensional (2D) spatial coordinate $\bm r = (x,y)$. The topological configurations of the field $\bm S(\bm r)$ shown in Fig.~\ref{fig:skyrmion}(a) and (b) are referred to as skyrmions.  Depending on a specific FM material, two distinct types of skyrmions are observed in experiment: the N\'eel (hedgehog) skyrmion and Bloch (spiral) skyrmion shown in Fig.~\ref{fig:skyrmion}(a) and (b), respectively. Although, the two types of skyrmions differ significantly in the orientation of the in-plane spins both are characterized by the same topological charge
\begin{align}
	Q = \frac{1}{4\pi} \int d^2r \, \hat {\bm S}\cdot (\nabla_x\hat {\bm S}\times\nabla_y\hat {\bm S})=1,\quad  \hat {\bm S}= \frac{\bm S}{S}.
	\label{topCharge}
\end{align}
Thus, one can transform a Neel skyrmion into a Bloch skyrmion by a $\pi/2$ rotation~\footnote{Note that for the case of a spin-singlet SC given by Eq.~(\ref{ham}), the Bloch and the Neel skyrmions are equivalent since they can be related by a continuous $\pi/2$-rotation around the $z$-axis in the spin space $U = \exp(-i\pi\sigma_z/4)$. In the presence of either the spin-triplet pairing or the spin-orbit interaction, the effects of the two types of skyrmions are different.} of the FM vector around the $\hat {\bm z}$ axis in the spin space without a change in the topological charge~(\ref{topCharge}).

Let us consider a heterostructure of a SC and FM with a skyrmion as shown in Fig.~\ref{fig:skyrmion}(a) and (b). The SC is described by the 4-by-4 Bogoliubov-de Gennes (BdG) Hamiltonian
\begin{align}
 H &= \xi(\bm p)\tau_z+\Delta \tau_x - \bm S(\bm r)\cdot\bm\sigma, \label{ham} \\
   & \xi(\bm p) = \frac{\bm p^2}{2m}-\mu,\quad \bm p = -i(\nabla_x,\nabla_y).
\end{align}
Here, $\xi(\bm p)$ describes the kinetic energy and $\Delta$ - the self-consistent superconducting gap, which we assume uniform in space; the term $\bm S(\bm r)\cdot\bm\sigma$ describes the proximity coupling between the FM film and SC. We assume that the Zeeman splitting $S(\bm r)$ does not exceed the Chandrasekhar-Clogston limit and $S<\Delta$. We also neglect the possible orbital effect of the magnetic field onto the superconductor \footnote{Neglecting the orbital effect of the magnetic field is reasonable for ferromagnetic films of atomic thickness, for example produced in Hamburg~\cite{Romming2013,Bergmann2014,Romming2015}. On the other hand, the orbital effect of the magnetic field cannot be neglected for the 3D ferromagnets. In this case, the magnetic field produces vortices in a type-II superconductor. Note that interaction between a skyrmion and vortex was recently considered in Ref. \cite{Hals2016}. }. The Pauli matrices $\bm \tau$ and $\bm \sigma$ act, respectively, in the particle-hole and spin subspaces of the four-component spinor $\Psi = (\psi_\uparrow,\psi_\downarrow,\psi^\dagger_\downarrow,-\psi^\dagger_\uparrow)^T$. At this point, we do not include the effects of the spin-orbit coupling or spin-triplet superconductivity~\cite{Note1} in the model~(\ref{ham}). We consider a case with a single N\'eel skyrmion centered at the origin, i.e. at $\bm r = 0$, and, so, assume the following profile of the FM vector
\begin{align}
	\bm S(\bm r) &= S\left[ \cos\phi(\bm r) \sin\theta(\bm r),\, \sin\phi(\bm r)\sin\theta(\bm r),\,\cos\theta(\bm r)\right],\nonumber  \\
	\phi(\bm r) &= \arctan(y/x),\quad \theta(\bm r) = \pi \left[ 1-\exp\left( -\frac{r^2}{R^2} \right) \right], \label{conf}
\end{align}
where $R$ defines an effective radius of the skyrmion \footnote{We expect that a different spatial dependency of the azimuthal angle (\ref{conf}) will not change the results significantly.}. Let us compare the relevant spatial scales of the problem: the SC coherence length $\xi_{\rm sc} \approx v_F/\Delta$, the skyrmion radius $R$, and the Fermi length $p_F^{-1}$. Both the scales $\xi_{\rm sc}$ and $R$ can vary from tens of nanometers to a micron depending on a specific material, whereas the Fermi length $p_F^{-1}$ is typically smaller than the other two scales. In the regime $R\gg \xi_{\rm sc}$, the skyrmion can be viewed as a large FM domain pointing in the direction opposite to the rest of the system. Such a regime could be interesting in the context of topological SC~\cite{Alicea2012}. For instance, it was recently shown \cite{Braunecker2013,Klinovaja2013,Vazifeh2013} that a helical texture of spins in a one-dimensional (1D) chain of magnetic atoms on a surface of a SC generates an effective Rashba-like spin-orbit interaction responsible for the Majorana edge modes. Similar effective spin-orbit interaction is generated near a skyrmion and could give rise to non-trivial edge states localized at the edge of the skyrmion. We leave the discussion of this  case for future works \footnote{Indeed, after the submission of our manuscript, Reference~\cite{Yang2016} found a zero energy Majorana bound state at the core of the skyrmion of a higher winding number.}. In the current paper, we focus on the case of relatively small skyrmions, i.e. $R\lesssim \xi_{sc}$.

\section{Multipole expansion of the skyrmion texture} \label{sec:analytics}
Let us first consider the case of a small skyrmion, i.e. $R\ll\xi_{\rm sc}$. In this limit, the superconductivity cannot ``resolve'' the fine details of the field $\bm S(\bm r)$. We perform the multipole expansion of the skyrmion configuration~(\ref{conf}) and approximate it as a point magnetic moment floating in a ``ferromagnetic sea'' as illustrated in Fig.~\ref{fig:skyrmion}(c)
\begin{align}
	\bm S_{\rm approx}(\bm r) & =  - S \hat{\bm z} + S_0 \hat{\bm z} \delta^2(\bm r),  \label{vr}
\end{align}
where $S_0$ is the zeroth moment of $\bm S(\bm r)$
\begin{align}
	S_0 = \int  d^2r \, \left[\bm S(\bm r)-\bm S(\infty)\right]_z  \quad \sim \,\,\,SR^2. \label{S0}
\end{align}
The formal domain of validity of the multipole expansion is $R \lesssim p_F^{-1} \ll \xi_{sc}$~\footnote{The domain of applicability can also be extended to $p_F^{-1}<R\ll \xi_{sc}$ with some modification of the theory.}. The multipole expansion gives an elegant and physically transparent description of the system, and, for this reason, we use it even beyond the domain of validity. In the end of the paper, we present an exact numerical modeling and find a close agreement with a multipole analytical treatment.

\begin{figure} \centering
	\includegraphics[width=0.7\linewidth]{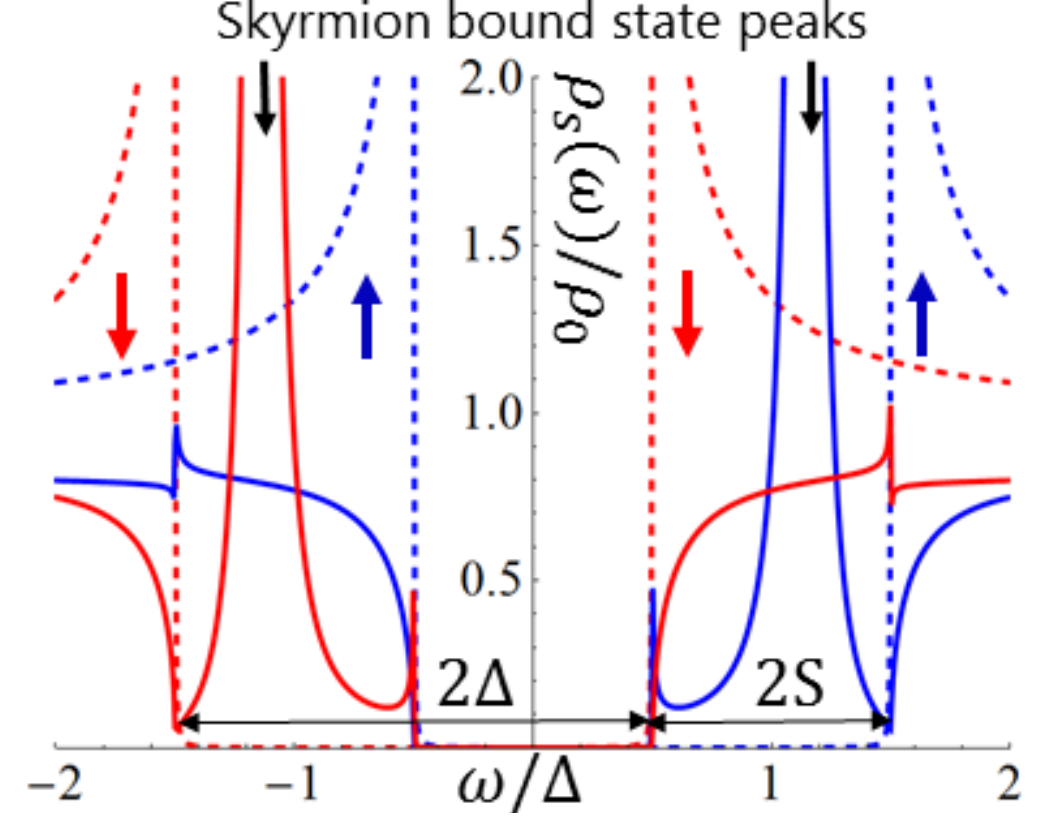}
	\caption{(color online) Spin-polarized local density of states (SP LDOS) of SC away from the skyrmion (dashed) and at the skyrmion core (solid). The color of the curves encodes the spin polarization: blue for spin up and red for spin down as indicated by the arrows. The figure is obtained by using a model given by Eqs.~(\ref{tm1}) and (\ref{ldos}) for the parameters $2S = \Delta = 0.1 \mu$, $R = 2.5/p_F$, $S_0 = 5SR^2$, and $S_1 = 0.5 SR^3$.} \label{fig:LDOS}
\end{figure}

By performing the T-matrix calculation, we solve the model given by Eqs.~(\ref{ham}) and (\ref{vr}), where we treat the local term $S_0 \hat{\bm z} \delta^2(\bm r)$ as a perturbation. We include the constant background magnetization $-S\hat{\bm z}$ in the BdG Hamiltonian $h(\bm p) = \xi(\bm p)\tau_z+\Delta \tau_x +  S\sigma_z$ and calculate an on-site matrix element of the bare Green's function $g(\omega,\bm p) = [\omega-h(\bm p)]^{-1}$
 \begin{align}
	 g_{0}(\omega)  =-\pi\rho_0\sum_{\lambda = \pm 1} \frac{1+\lambda\sigma_z}{2}\,\frac{\omega-\lambda S+\Delta\tau_x}{\sqrt{\Delta^2-\left( \omega-\lambda S \right)^2}},  \label{grf}
\end{align}
where $\rho = m/2\pi$ is the density of states. This Green's function describes a SC subject to a uniform background magnetization $-\hat{\bm z} S$ that shifts the spin subbands as shown with the dashed lines in Fig.~\ref{fig:LDOS}.  The density of states contains two interior and two exterior coherence peaks at the energies $\pm(\Delta-S)$ and $\pm(\Delta+S)$ correspondingly.  Using Green's function~(\ref{grf}) we calculate the T-matrix in the presence of a point magnetic moment $V(\bm r)=-S_0\,\sigma_z \delta^2(\bm r)$ representing the skyrmion
\begin{align}
	T(\omega) =   \frac{-S_0\sigma_z}{1+S_0\sigma_zg_{0}(\omega)}. \label{tm}
\end{align}
The poles of T-matrix give the energies of the skyrmion-induced bound states 
\begin{align}
	E^\pm_{\rm SBS} = \pm\left[S+\Delta \frac{1-\left( \pi\rho S_0 \right)^2}{1+\left( \pi\rho S_0 \right)^2}\right].
	\label{energy}
\end{align}
Let us trace the bound state energies as a function of increasing $S_0$, which is an implicit function of $S$ and $R$ according to Eq.~(\ref{S0}). For small $S_0$, the bound states lie at the outer coherence peaks at the energies $\pm (\Delta+S)$. With further increase of $S_0$, the bound states energies split from the outer coherence peak and move to the inner coherence peaks \footnote{Although Eq.~(\ref{energy}) suggests that the bound states may go inside the actual gap for large enough  $S_0$, i.e. $|E^\pm_{\rm SBS}|<\Delta-S$, the multipole approximation of a skyrmion (\ref{vr}) is no-longer valid in this regime and, thus, does not give a reliable estimate of the energy. In practice, by performing a numerical modeling, we never observe the bound state peaks inside the actual spectral gap, i.e. in the window of energies $|E^\pm_{\rm SBS}|<\Delta-S$.}. The spin-polarization of the bound states is determined by the spin-polarization of the bulk bands that they split from: the positive (negative) state is ``up'' (``down'') spin-polarized. The bound states closely resemble the well-known Yu-Shiba-Rusinov (YSR) states \cite{Yu,Shiba,Rusinov,Balatsky2006} localized around magnetic impurities in SC. The main difference is that the YSR energies reside inside the actual spectral gap, whereas the bound states energies lie in the window of energies $\Delta+S>|E^\pm_{\rm SBS}|>\Delta-S$, which is also filled with a continuum of delocalized states of the opposite spin polarization.

Now let us show that the bound states give resonances of finite spectral width due to the coupling with the continuum of delocalized states. Indeed, the skyrmion has in-plane spins at $r\approx R$ that couple the spin-up and spin-down sectors of the Hamiltonian. In order to capture this effect we append the multipole expansion (\ref{vr}) with a next order term  representing the radial in-plane spins of the skyrmion.
\begin{align}
	\bm S_{\rm approx}(\bm r) & =  - S \hat{\bm z} + S_0 \hat{\bm z} \delta^2(\bm r)-S_1 \bm \nabla \delta^2(\bm r), \label{vr1}
\end{align}
where  $\bm \nabla = (\nabla_x,\nabla_y)$ and $S_1$ is the first moment of the original skyrmion configuration $\bm S(\bm r)$
\begin{align}
	S_1 = \frac{1}{2}\int  d^2r \, \left[\bm S(\bm r)-\bm S(\infty)\right] \cdot \bm r\quad \sim \,\,\,SR^3. \label{S1}
\end{align}
In Appendix~\ref{sec:appendixTMatrix}, we solve the Lippmann-Schwinger equation for the T-matrix for Eqs.~(\ref{ham}) and (\ref{vr1}) 
\begin{align}
	T(\omega) =   \frac{-S_0\sigma_z+S^2_1p_F^2\bar g_{0}(\omega)}{1+S_0\sigma_zg_{0}(\omega)-S^2_1p_F^2\,\bar g_{0}(\omega)\, g_{0}(\omega)}. \label{tm1}
\end{align}
Here, the Green's function $\bar g_0(\omega) = \frac{1}{2}\sum_{j=x,y} \sigma_j g_0(\omega) \sigma_j $ describes the bands with opposite spin polarization ${\sigma_z\rightarrow -\sigma_z}$. Using Eq.~(\ref{tm1}) we calculate SP LDOS
\begin{align}
	\rho_s & (\omega) = \label{ldos} \\
	&-\frac{1}{\pi}\,{\rm Im}{\rm \,Tr} \left\{  \frac{1+\tau_z}{2}\,\frac{1+\sigma_s}{2} \left[g_0(\omega)+g_0(\omega) T(\omega) g_0(\omega)  \right]\right\},\nonumber
\end{align}
where $s=x,y,z$ denotes the spin projection axis. We plot the LDOS (\ref{ldos}) with solid lines in Fig.~\ref{fig:LDOS} and compare it with LDOS away from the skyrmion shown with dashed lines. We observe that the peaks corresponding to the bound states have finite spectral width. Indeed, the denominator of T-matrix~(\ref{tm1}) has an extra term compared to that of Eq.~(\ref{tm}). The first two terms in the denominator of (\ref{tm1}) give the bound states energies~(\ref{energy}), whereas the last term $S^2_1p_F^2\,\bar g_{0}(\omega)\, g_{0}(\omega)$ is imaginary and defines the spectral width of the resonances observed in Fig.~\ref{fig:LDOS}.

\begin{figure} \centering
	(a)\includegraphics[width=0.65\linewidth]{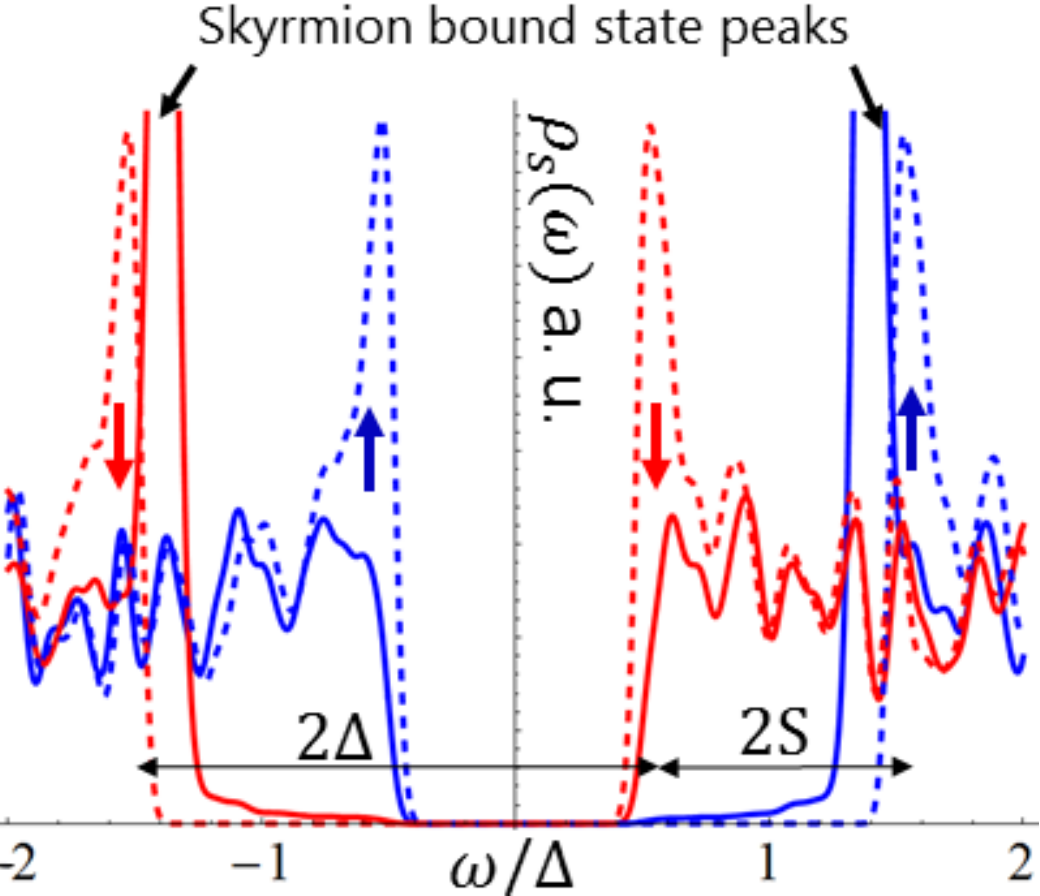} \\
	(b)\includegraphics[width=0.65\linewidth]{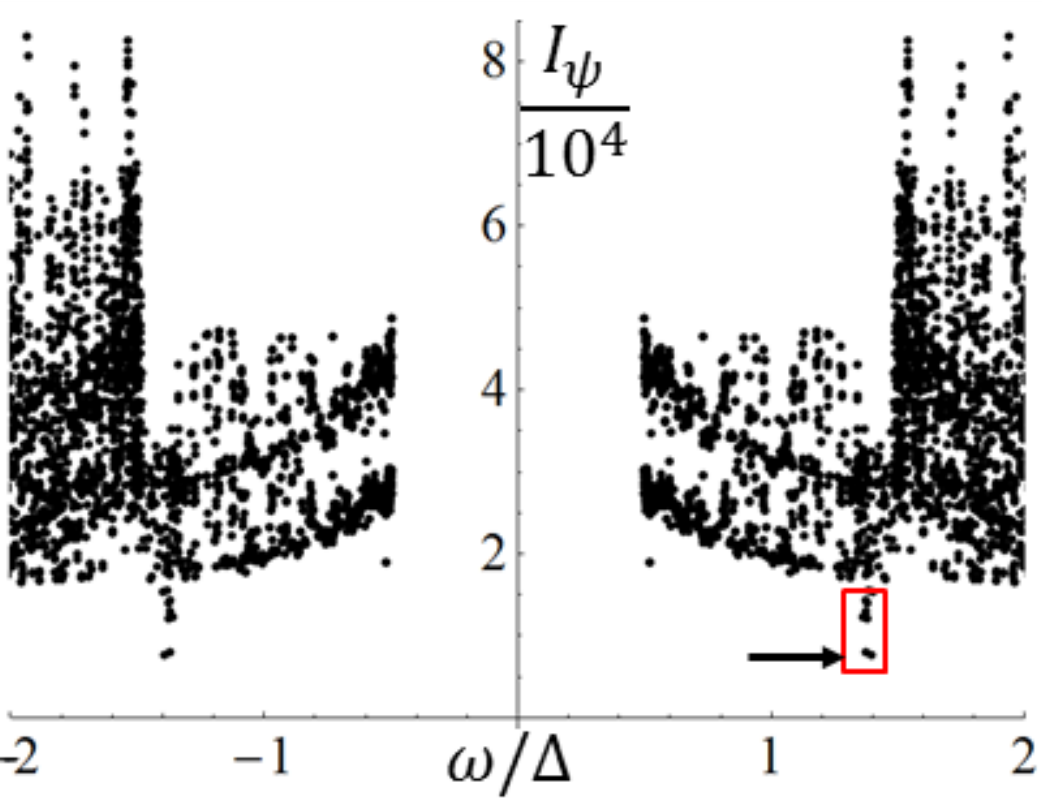}
	\caption{(color online) Numerical modeling of a skyrmion. (a) Spin-polarized LDOS at the skyrmion core (solid) and away from the skyrmion (dashed).  (b) The function $I_\psi$ characterizing a degree of localization of each BdG wavefunctions $\psi$ versus eigenenergy $\omega$. Few of the quasilocalized wavefunctions emphasized by the red rectangle are shown in Fig.~\ref{fig:wavefunction}.} \label{fig:LDOSNumerics}
\end{figure}

\section{Numerical analysis} \label{sec:numerics}

\begin{figure} \centering
	(a)\includegraphics[width=0.43\linewidth]{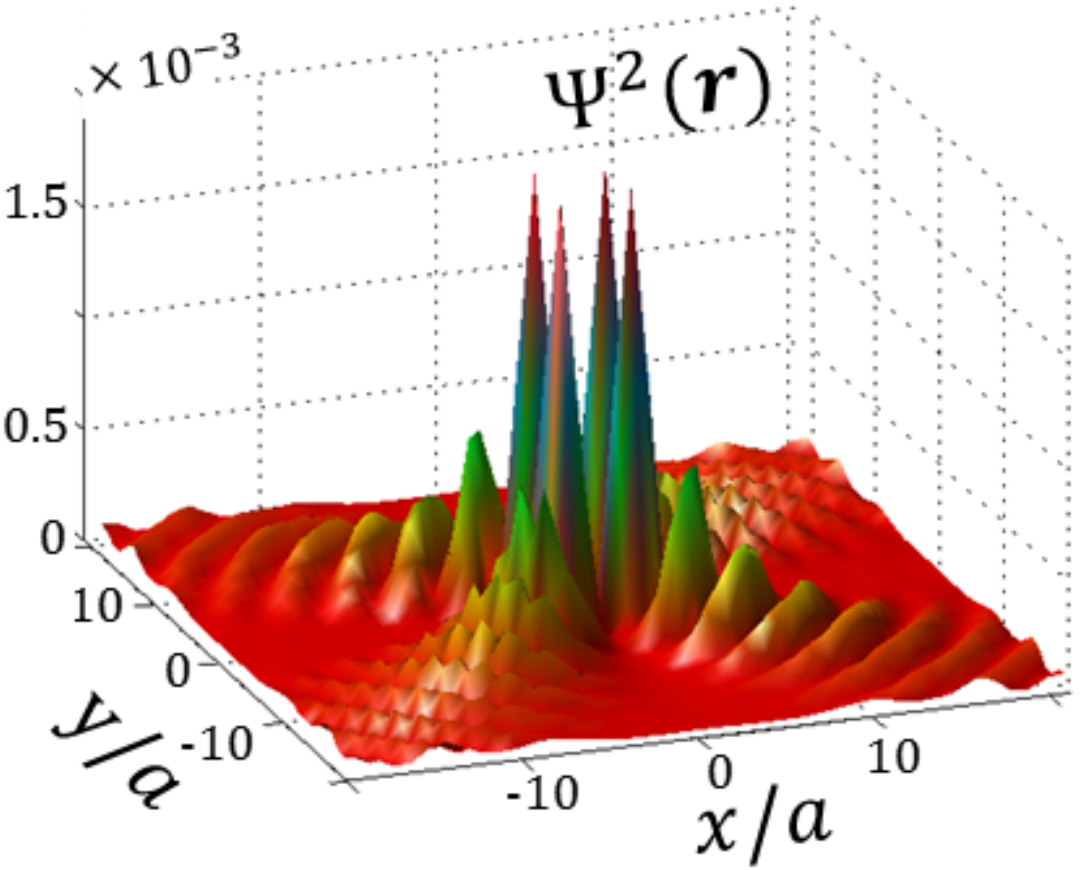}
	(b)\includegraphics[width=0.43\linewidth]{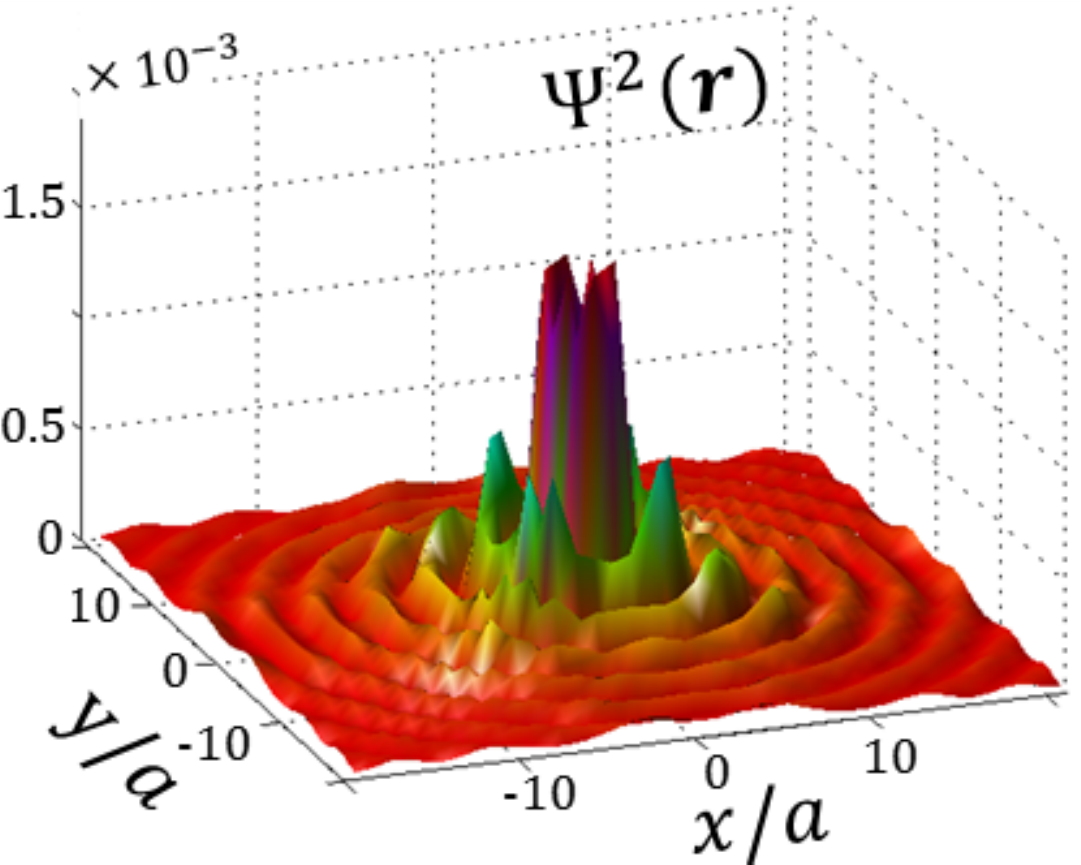} \\
	(c)\includegraphics[width=0.43\linewidth]{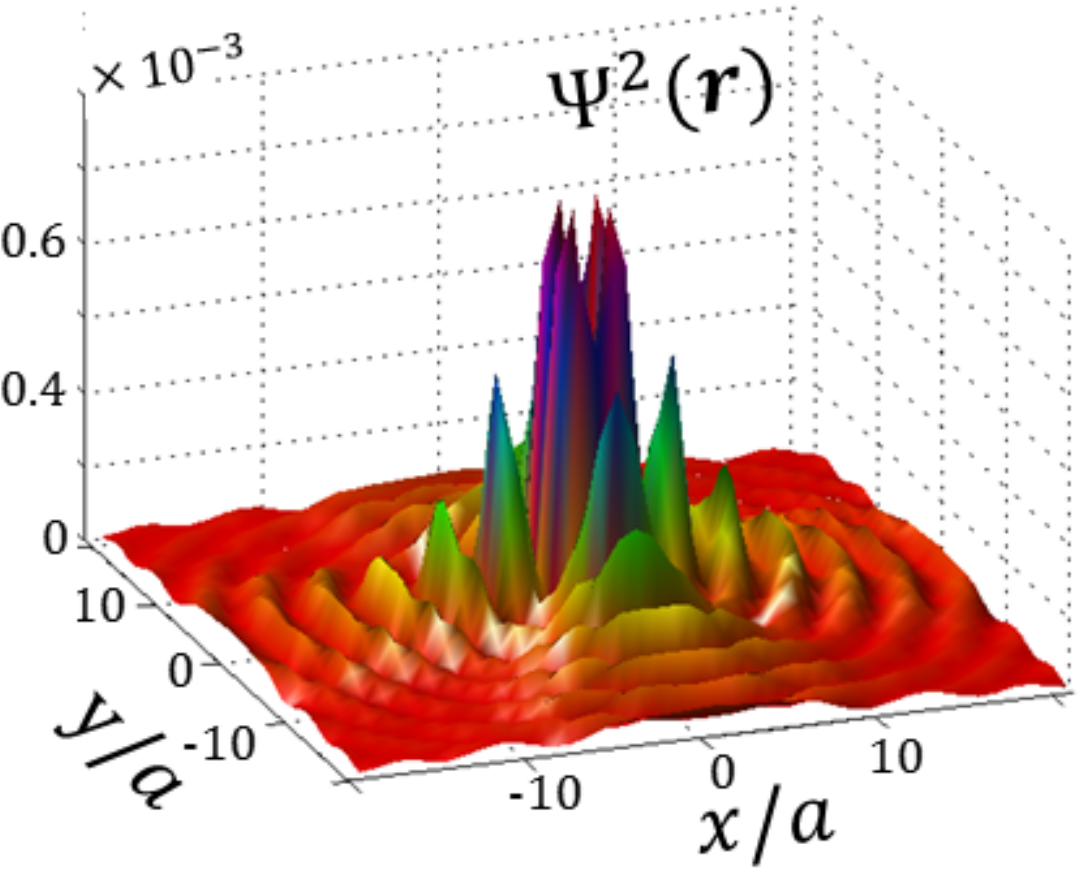}
	(d)\includegraphics[width=0.43\linewidth]{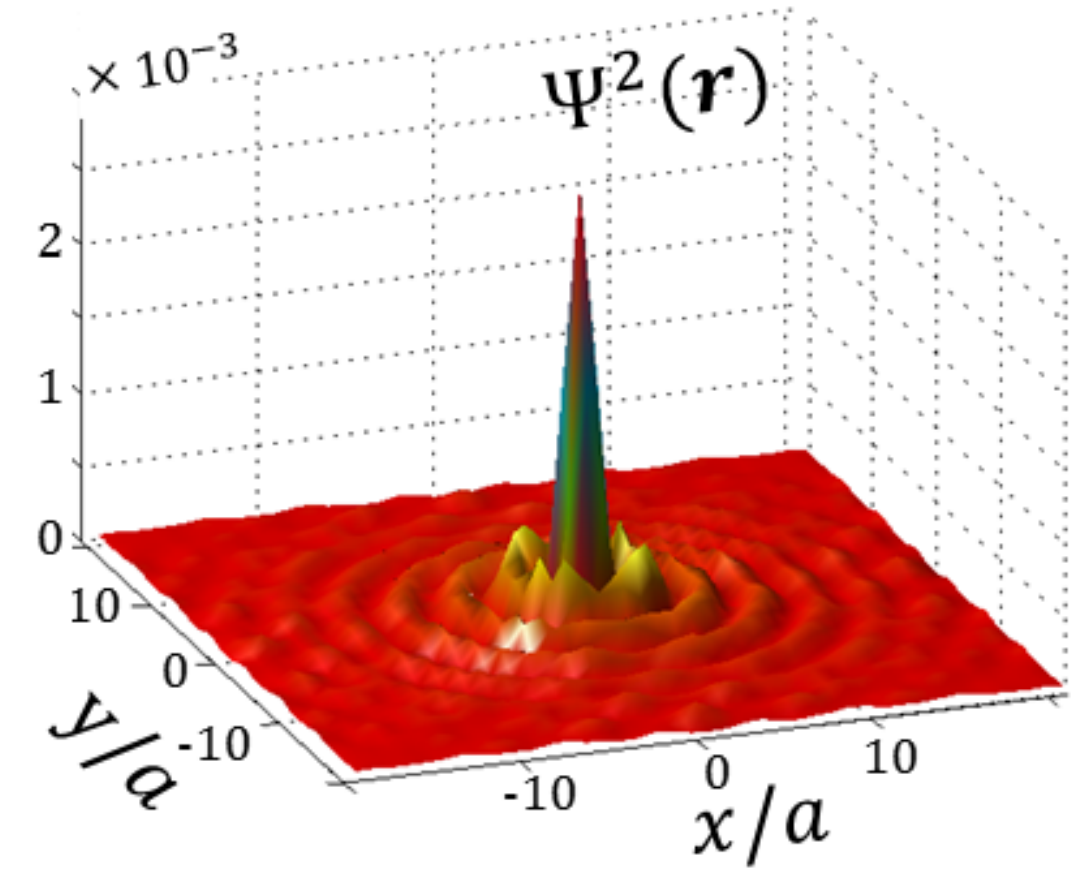}
	 \caption{(Color online) Spatial profile of the quasilocalized wavefunctions obtained numerically, which are indicated in Fig.~\ref{fig:LDOSNumerics}(b). The wavefunctions are shown in the order of increasing $I_\psi$.  } \label{fig:wavefunction}
\end{figure}

So far we have analyzed the skyrmion using the analytical multipole approximation. Now let us present the results of an exact numerical modeling. We set the BdG Hamiltonian on the N-by-N tight-binding square lattice with parameters: the lateral size of the system N = 200, nearest neighbor coupling $t$, on-site superconducting pairing and Zeeman coupling $\Delta=2S=0.1t$, and chemical potential $\mu = -3t$. This choice of parameters corresponds to $\xi_{\rm sc}\approx 17 a$ in the units of the elementary cell constant $a$. The skyrmion is described by the vector $\bm S(\bm r)$ given by Eq.~(\ref{conf}) with the effective radius $R = 6a$, so that $2R\sim \xi_{\rm sc}$. From the numerical wavefunctions, we calculate SP LDOS, apply the Gaussian smoothing kernel and plot the resulting SP LDOS in Fig.~\ref{fig:LDOSNumerics}(a). We use the same plotting style as in Fig.~\ref{fig:LDOS}: solid (dashed) line represents LDOS at (away from) the skyrmion, whereas colors encode spin polarizations. We observe that the calculated LDOS is consistent with the results of the analytical calculation. Away from the skyrmion, SP LDOS contains the shifted spin subbands. At the skyrmion core, the skyrmion induces a strong resonance in the energy window $\Delta-S<|\omega|<\Delta+S$. In order to further analyze the numerical wavefunctions $\psi(\bm r)$, we also calculate the following expression
\begin{equation}
	I_\psi =\frac{1}{\sum_{\bm r,j} |\psi_j(\bm r)|^4},
	\label{I}
\end{equation}
where the sum is carried over all lattice sites $\bm r$ as well as all components $j = 1,\ldots,4$ of the four-component BdG wavefunction on each site. The function $I_\psi$ characterizes a degree of a localization of the wavefunction $\psi(\bm r)$ \footnote{The function $I_\psi^{-1}$ is commonly referred to as the inverse participation ratio in the literature on localization phenomena.}. The function is small $I_\psi \sim 1$ for an extremely localized wavefunction and large $I_\psi \sim N^2$ for a delocalized wavefunction. For each numerical BdG wavefunction $\psi(\bm r)$, we plot a map of $I_\psi$ versus the eigenenergy in Fig.~\ref{fig:LDOSNumerics}(b). We observe a number of distinct quasilocalized states that stand out from the rest of states as emphasized by the red rectangle in Fig.~\ref{fig:LDOSNumerics}(b). These states have the energy of the bound states peak. The number of the quasilocalized states grows with skyrmion size. We visualize the spatial profile of the electron part of the BdG wavefunction $|\Psi(\bm r)|^2 = |u_{\uparrow}(\bm r)|^2+|u_{\downarrow}(\bm r)|^2$  for a few of these states in Fig.~\ref{fig:wavefunction}. In contrast with the analytical results, we find a wavefunction with multiple lobes corresponding to a higher angular momentum state, shown in panel (a), as well as a state with a single peak, shown in panel (d). It is known that higher angular momentum states do form bound states. Analytic solution presented above is based on the on-site T matrix and is not sufficient to capture the higher-angular-momentum bound states.

We also observe that all wavefunctions in Fig.~\ref{fig:wavefunction} exhibit characteristic oscillations at the scale $\xi_{\rm sc}$. In order to understand this behavior, consider a generic wavefunction of an impurity induced state ${\Psi_\lambda(\bm r) \sim e^{ip_Fr-r\sqrt{\Delta^2-(\omega - \lambda S)^2}/v_F}/\sqrt{r}}$, where $\lambda$ denotes the eigenvalues of $\sigma_z$ operator. The terms in the exponent term describe behavior at two scales $p_F^{-1}$ as well as $\xi_{\rm sc}$. For clarity, let us focus on the positive bound state, i.e. $\omega = E^{+}_{\rm SBS}$. From the point of view of spin-up subband, i.e. $\lambda = 1$, the $E^{+}_{\rm SBS}$ state is subgap, i.e. $|\omega- S|<\Delta$, and so the square root term in $\Psi_+(\bm r)$ gives an exponentially localized wavefunction. However from the point of view of spin-down subband, i.e. $\lambda = -1$, the $E^{-}_{\rm SBS}$ state is supragap, i.e. $\omega+S>\Delta$, and, so, the square root in $\Psi_-(\bm r)$ gives oscillations  at the scale of $\xi_{sc}$ superimposed with a long-range $1/\sqrt{r}$ decay. These oscillations as well the long-range behavior can be seen in Fig.~\ref{fig:wavefunction}.

\section{Interaction between skyrmions mediated by a superconductor.} \label{sec:interaction}

 Reference~\cite{Menard2015} reported the STM study of the YSR states induced by the magnetic dopants in a quasi-2D superconductor. In contrast with the previous experiments, which observed YSR states only on the atomic scale, it was demonstrated the YSR wavefunction can extend over the range of tens of nanometers, i.e. two orders of magnitude greater than observed before. Theoretical paper~\cite{Yao2014} argued that superconductivity induces an effective interaction between the magnetic spins when the corresponding YSR wavefunctions overlap. Thus, motivated by Refs.~\cite{Menard2015,Yao2014}, we propose that superconductiviy could mediate an effective interaction between the distinct magnetic skyrmions when the corresponding long-range skyrmion bounds states overlap. 
 
 Let us briefly sketch the argument given in Appendix~\ref{sec:appendixInteraction}, where a perturbative in $S_1$ derivation of the skyrmion-skyrmion interaction is given. Consider two skyrmions in the ferromagnetic film proximity coupled to a superconductor. As was shown in the previous sections, each skyrmion induces a spin-polarized resonance in the window of energies $\Delta - S <|E_{\rm SBS}|<\Delta+S$ also populated by the delocalized states of the opposite spin-polarization. In the limit $S_1=0$, the bound state wavefunctions, being subgap states of the corresponding spin-polarized sector of the Hamiltonian, are exponentially localized. If $S_1\neq 0$, the corresponding perturbation $S_1\, \left(\bm \sigma \cdot  \bm \nabla\right) \delta^2(\bm r)$ contains in-plane Pauli matrices $(\sigma_x,\sigma_y)$, which mix the opposite spin sectors of the Hamiltonian. Therefore, the skyrmion bound states corresponding to the distinct skyrmions can couple and hybridize via the long-range delocalized states of the opposite spin-polarization. This will have an energetic effect leading to an effective long-range interaction between the skyrmions.

\section{Conclusion and Outlook.} \label{sec:conclusion}

In this paper, we predict the new skyrmion bound states in the superconductor proximity coupled to the ferromagnetic film with a skyrmion texture. We calculate spin-polarized local density of states and show the signatures of the bound states in the tunneling spectrum that could be measured by the spin-polarized scanning tunneling microscopy. By using an analogy with the well-known YSR states, we show that the skyrmion induces a resonance in between the spin-split coherence peaks corresponding to the opposite spin polarizations. We show that the corresponding wavefunction is long-range in contrast with the YSR states, which are short-range. Thus in the case of the two skyrmions, the corresponding wavefunctions will overlap and induce a long-range interaction between the skyrmions \cite{Yao2014,Menard2015}. 


After this manuscript was submitted, we learned about further theoretical studies \cite{Yang2016,Hals2016} of the hybrid skyrmion-SC heterostructures. Authors of Ref.~\cite{Yang2016} found the Majorana bound state solution in the vicinity of the skyrmions of higher winding numbers. Authors of Ref.~\cite{Hals2016} studied an interaction between a skyrmion and vortex in a type-II superconductor. 

We thank  R.~Wiesendanger, S.~Fujimoto, J.~Wiebe, J.~Zang, A.~Saxena, H. Hurst, Y. Tserkovnyak, S. Lin and L. Bulaevskii for valuable discussions and comments. This work was supported by US DOE BES E304 (S.S.P. and A.V.B.) and by the Grant-in-Aid for Research Activity Start-up (No.~15H06858) (S.N.).

\bibliography{Skyrmion}

\begin{thebibliography}{44}%
\makeatletter
\providecommand \@ifxundefined [1]{%
 \@ifx{#1\undefined}
}%
\providecommand \@ifnum [1]{%
 \ifnum #1\expandafter \@firstoftwo
 \else \expandafter \@secondoftwo
 \fi
}%
\providecommand \@ifx [1]{%
 \ifx #1\expandafter \@firstoftwo
 \else \expandafter \@secondoftwo
 \fi
}%
\providecommand \natexlab [1]{#1}%
\providecommand \enquote  [1]{``#1''}%
\providecommand \bibnamefont  [1]{#1}%
\providecommand \bibfnamefont [1]{#1}%
\providecommand \citenamefont [1]{#1}%
\providecommand \href@noop [0]{\@secondoftwo}%
\providecommand \href [0]{\begingroup \@sanitize@url \@href}%
\providecommand \@href[1]{\@@startlink{#1}\@@href}%
\providecommand \@@href[1]{\endgroup#1\@@endlink}%
\providecommand \@sanitize@url [0]{\catcode `\\12\catcode `\$12\catcode
  `\&12\catcode `\#12\catcode `\^12\catcode `\_12\catcode `\%12\relax}%
\providecommand \@@startlink[1]{}%
\providecommand \@@endlink[0]{}%
\providecommand \url  [0]{\begingroup\@sanitize@url \@url }%
\providecommand \@url [1]{\endgroup\@href {#1}{\urlprefix }}%
\providecommand \urlprefix  [0]{URL }%
\providecommand \Eprint [0]{\href }%
\providecommand \doibase [0]{http://dx.doi.org/}%
\providecommand \selectlanguage [0]{\@gobble}%
\providecommand \bibinfo  [0]{\@secondoftwo}%
\providecommand \bibfield  [0]{\@secondoftwo}%
\providecommand \translation [1]{[#1]}%
\providecommand \BibitemOpen [0]{}%
\providecommand \bibitemStop [0]{}%
\providecommand \bibitemNoStop [0]{.\EOS\space}%
\providecommand \EOS [0]{\spacefactor3000\relax}%
\providecommand \BibitemShut  [1]{\csname bibitem#1\endcsname}%
\let\auto@bib@innerbib\@empty
\bibitem [{\citenamefont {Skyrme}(1961)}]{Skyrme}%
  \BibitemOpen
  \bibfield  {author} {\bibinfo {author} {\bibfnamefont {T.~H.}\ \bibnamefont
  {Skyrme}},\ }\bibfield  {title} {\enquote {\bibinfo {title} {A non-linear
  field theory},}\ }\href@noop {} {\bibfield  {journal} {\bibinfo  {journal}
  {Proc. R. Soc. Lond. Ser. A}\ }\textbf {\bibinfo {volume} {260}},\ \bibinfo
  {pages} {127} (\bibinfo {year} {1961})}\BibitemShut {NoStop}%
\bibitem [{\citenamefont {Bogdanov}\ and\ \citenamefont
  {Yablonskii}(1989)}]{Bogdanov1989}%
  \BibitemOpen
  \bibfield  {author} {\bibinfo {author} {\bibfnamefont {A.~N.}\ \bibnamefont
  {Bogdanov}}\ and\ \bibinfo {author} {\bibfnamefont {D.~A.}\ \bibnamefont
  {Yablonskii}},\ }\bibfield  {title} {\enquote {\bibinfo {title}
  {{Thermodynamically stable "vortices" in magnetically ordered crystals. The
  mixed state of magnets.}}}\ }\href
  {http://www.jetp.ac.ru/cgi-bin/e/index/e/68/1/p101?a=list} {\bibfield
  {journal} {\bibinfo  {journal} {Sov. Phys. JETP}\ }\textbf {\bibinfo {volume}
  {68}},\ \bibinfo {pages} {101} (\bibinfo {year} {1989})}\BibitemShut
  {NoStop}%
\bibitem [{\citenamefont {R\"{o}ssler}\ \emph {et~al.}(2006)\citenamefont
  {R\"{o}ssler}, \citenamefont {Bogdanov},\ and\ \citenamefont
  {Pfleiderer}}]{Rossler2006}%
  \BibitemOpen
  \bibfield  {author} {\bibinfo {author} {\bibfnamefont {U.~K.}\ \bibnamefont
  {R\"{o}ssler}}, \bibinfo {author} {\bibfnamefont {A.~N.}\ \bibnamefont
  {Bogdanov}}, \ and\ \bibinfo {author} {\bibfnamefont {C.}~\bibnamefont
  {Pfleiderer}},\ }\bibfield  {title} {\enquote {\bibinfo {title} {{Spontaneous
  skyrmion ground states in magnetic metals.}}}\ }\href {\doibase
  10.1038/nature05056} {\bibfield  {journal} {\bibinfo  {journal} {Nature}\
  }\textbf {\bibinfo {volume} {442}},\ \bibinfo {pages} {797} (\bibinfo {year}
  {2006})}\BibitemShut {NoStop}%
\bibitem [{\citenamefont {M\"{u}hlbauer}\ \emph {et~al.}(2009)\citenamefont
  {M\"{u}hlbauer}, \citenamefont {Binz}, \citenamefont {Jonietz}, \citenamefont
  {Pfleiderer}, \citenamefont {Rosch}, \citenamefont {Neubauer}, \citenamefont
  {Georgii},\ and\ \citenamefont {B\"{o}ni}}]{Muhlbauer2009}%
  \BibitemOpen
  \bibfield  {author} {\bibinfo {author} {\bibfnamefont {S.}~\bibnamefont
  {M\"{u}hlbauer}}, \bibinfo {author} {\bibfnamefont {B.}~\bibnamefont {Binz}},
  \bibinfo {author} {\bibfnamefont {F.}~\bibnamefont {Jonietz}}, \bibinfo
  {author} {\bibfnamefont {C.}~\bibnamefont {Pfleiderer}}, \bibinfo {author}
  {\bibfnamefont {A.}~\bibnamefont {Rosch}}, \bibinfo {author} {\bibfnamefont
  {A.}~\bibnamefont {Neubauer}}, \bibinfo {author} {\bibfnamefont
  {R.}~\bibnamefont {Georgii}}, \ and\ \bibinfo {author} {\bibfnamefont
  {P.}~\bibnamefont {B\"{o}ni}},\ }\bibfield  {title} {\enquote {\bibinfo
  {title} {{Skyrmion lattice in a chiral magnet.}}}\ }\href {\doibase
  10.1126/science.1166767} {\bibfield  {journal} {\bibinfo  {journal}
  {Science}\ }\textbf {\bibinfo {volume} {323}},\ \bibinfo {pages} {915}
  (\bibinfo {year} {2009})}\BibitemShut {NoStop}%
\bibitem [{\citenamefont {M\"unzer}\ \emph {et~al.}(2010)\citenamefont
  {M\"unzer}, \citenamefont {Neubauer}, \citenamefont {Adams}, \citenamefont
  {M\"uhlbauer}, \citenamefont {Franz}, \citenamefont {Jonietz}, \citenamefont
  {Georgii}, \citenamefont {B\"oni}, \citenamefont {Pedersen}, \citenamefont
  {Schmidt}, \citenamefont {Rosch},\ and\ \citenamefont
  {Pfleiderer}}]{Munzer2010}%
  \BibitemOpen
  \bibfield  {author} {\bibinfo {author} {\bibfnamefont {W.}~\bibnamefont
  {M\"unzer}}, \bibinfo {author} {\bibfnamefont {A.}~\bibnamefont {Neubauer}},
  \bibinfo {author} {\bibfnamefont {T.}~\bibnamefont {Adams}}, \bibinfo
  {author} {\bibfnamefont {S.}~\bibnamefont {M\"uhlbauer}}, \bibinfo {author}
  {\bibfnamefont {C.}~\bibnamefont {Franz}}, \bibinfo {author} {\bibfnamefont
  {F.}~\bibnamefont {Jonietz}}, \bibinfo {author} {\bibfnamefont
  {R.}~\bibnamefont {Georgii}}, \bibinfo {author} {\bibfnamefont
  {P.}~\bibnamefont {B\"oni}}, \bibinfo {author} {\bibfnamefont
  {B.}~\bibnamefont {Pedersen}}, \bibinfo {author} {\bibfnamefont
  {M.}~\bibnamefont {Schmidt}}, \bibinfo {author} {\bibfnamefont
  {A.}~\bibnamefont {Rosch}}, \ and\ \bibinfo {author} {\bibfnamefont
  {C.}~\bibnamefont {Pfleiderer}},\ }\bibfield  {title} {\enquote {\bibinfo
  {title} {{Skyrmion lattice in the doped semiconductor
  ${\text{Fe}}_{1\ensuremath{-}x}{\text{Co}}_{x}\text{Si}$}},}\ }\href
  {\doibase 10.1103/PhysRevB.81.041203} {\bibfield  {journal} {\bibinfo
  {journal} {Phys. Rev. B}\ }\textbf {\bibinfo {volume} {81}},\ \bibinfo
  {pages} {041203} (\bibinfo {year} {2010})}\BibitemShut {NoStop}%
\bibitem [{\citenamefont {Yu}\ \emph {et~al.}(2011)\citenamefont {Yu},
  \citenamefont {Kanazawa}, \citenamefont {Onose}, \citenamefont {Kimoto},
  \citenamefont {Zhang}, \citenamefont {Ishiwata}, \citenamefont {Matsui},\
  and\ \citenamefont {Tokura}}]{Yu2011}%
  \BibitemOpen
  \bibfield  {author} {\bibinfo {author} {\bibfnamefont {X.~Z.}\ \bibnamefont
  {Yu}}, \bibinfo {author} {\bibfnamefont {N.}~\bibnamefont {Kanazawa}},
  \bibinfo {author} {\bibfnamefont {Y.}~\bibnamefont {Onose}}, \bibinfo
  {author} {\bibfnamefont {K.}~\bibnamefont {Kimoto}}, \bibinfo {author}
  {\bibfnamefont {W.~Z.}\ \bibnamefont {Zhang}}, \bibinfo {author}
  {\bibfnamefont {S.}~\bibnamefont {Ishiwata}}, \bibinfo {author}
  {\bibfnamefont {Y.}~\bibnamefont {Matsui}}, \ and\ \bibinfo {author}
  {\bibfnamefont {Y.}~\bibnamefont {Tokura}},\ }\bibfield  {title} {\enquote
  {\bibinfo {title} {{Near room-temperature formation of a skyrmion crystal in
  thin-films of the helimagnet FeGe}},}\ }\href {\doibase 10.1038/nmat2916}
  {\bibfield  {journal} {\bibinfo  {journal} {Nat. Mater.}\ }\textbf {\bibinfo
  {volume} {10}},\ \bibinfo {pages} {106} (\bibinfo {year} {2011})}\BibitemShut
  {NoStop}%
\bibitem [{\citenamefont {Heinze}\ \emph {et~al.}(2011)\citenamefont {Heinze},
  \citenamefont {von Bergmann}, \citenamefont {Menzel}, \citenamefont {Brede},
  \citenamefont {Kubetzka}, \citenamefont {Wiesendanger}, \citenamefont
  {Bihlmayer},\ and\ \citenamefont {Blugel}}]{Heinze2011}%
  \BibitemOpen
  \bibfield  {author} {\bibinfo {author} {\bibfnamefont {S.}~\bibnamefont
  {Heinze}}, \bibinfo {author} {\bibfnamefont {K.}~\bibnamefont {von
  Bergmann}}, \bibinfo {author} {\bibfnamefont {M.}~\bibnamefont {Menzel}},
  \bibinfo {author} {\bibfnamefont {J.}~\bibnamefont {Brede}}, \bibinfo
  {author} {\bibfnamefont {A.}~\bibnamefont {Kubetzka}}, \bibinfo {author}
  {\bibfnamefont {R.}~\bibnamefont {Wiesendanger}}, \bibinfo {author}
  {\bibfnamefont {G.}~\bibnamefont {Bihlmayer}}, \ and\ \bibinfo {author}
  {\bibfnamefont {S.}~\bibnamefont {Blugel}},\ }\bibfield  {title} {\enquote
  {\bibinfo {title} {{Spontaneous atomic-scale magnetic skyrmion lattice in two
  dimensions}},}\ }\href {\doibase 10.1038/nphys2045} {\bibfield  {journal}
  {\bibinfo  {journal} {Nat. Phys.}\ }\textbf {\bibinfo {volume} {7}},\
  \bibinfo {pages} {713--718} (\bibinfo {year} {2011})}\BibitemShut {NoStop}%
\bibitem [{\citenamefont {Seki}\ \emph {et~al.}(2012)\citenamefont {Seki},
  \citenamefont {Yu}, \citenamefont {Ishiwata},\ and\ \citenamefont
  {Tokura}}]{Seki2012}%
  \BibitemOpen
  \bibfield  {author} {\bibinfo {author} {\bibfnamefont {S.}~\bibnamefont
  {Seki}}, \bibinfo {author} {\bibfnamefont {X.~Z.}\ \bibnamefont {Yu}},
  \bibinfo {author} {\bibfnamefont {S.}~\bibnamefont {Ishiwata}}, \ and\
  \bibinfo {author} {\bibfnamefont {Y.}~\bibnamefont {Tokura}},\ }\bibfield
  {title} {\enquote {\bibinfo {title} {{Observation of Skyrmions in a
  Multiferroic Material}},}\ }\href {\doibase 10.1126/science.1214143}
  {\bibfield  {journal} {\bibinfo  {journal} {Science}\ }\textbf {\bibinfo
  {volume} {336}},\ \bibinfo {pages} {198} (\bibinfo {year}
  {2012})}\BibitemShut {NoStop}%
\bibitem [{\citenamefont {Jonietz}\ \emph {et~al.}(2010)\citenamefont
  {Jonietz}, \citenamefont {Mühlbauer}, \citenamefont {Pfleiderer},
  \citenamefont {Neubauer}, \citenamefont {Münzer}, \citenamefont {Bauer},
  \citenamefont {Adams}, \citenamefont {Georgii}, \citenamefont {Böni},
  \citenamefont {Duine}, \citenamefont {Everschor}, \citenamefont {Garst},\
  and\ \citenamefont {Rosch}}]{Jonietz2010}%
  \BibitemOpen
  \bibfield  {author} {\bibinfo {author} {\bibfnamefont {F.}~\bibnamefont
  {Jonietz}}, \bibinfo {author} {\bibfnamefont {S.}~\bibnamefont {Mühlbauer}},
  \bibinfo {author} {\bibfnamefont {C.}~\bibnamefont {Pfleiderer}}, \bibinfo
  {author} {\bibfnamefont {A.}~\bibnamefont {Neubauer}}, \bibinfo {author}
  {\bibfnamefont {W.}~\bibnamefont {Münzer}}, \bibinfo {author} {\bibfnamefont
  {A.}~\bibnamefont {Bauer}}, \bibinfo {author} {\bibfnamefont
  {T.}~\bibnamefont {Adams}}, \bibinfo {author} {\bibfnamefont
  {R.}~\bibnamefont {Georgii}}, \bibinfo {author} {\bibfnamefont
  {P.}~\bibnamefont {Böni}}, \bibinfo {author} {\bibfnamefont {R.~A.}\
  \bibnamefont {Duine}}, \bibinfo {author} {\bibfnamefont {K.}~\bibnamefont
  {Everschor}}, \bibinfo {author} {\bibfnamefont {M.}~\bibnamefont {Garst}}, \
  and\ \bibinfo {author} {\bibfnamefont {A.}~\bibnamefont {Rosch}},\ }\bibfield
   {title} {\enquote {\bibinfo {title} {{Spin Transfer Torques in MnSi at
  Ultralow Current Densities}},}\ }\href {\doibase 10.1126/science.1195709}
  {\bibfield  {journal} {\bibinfo  {journal} {Science}\ }\textbf {\bibinfo
  {volume} {330}},\ \bibinfo {pages} {1648} (\bibinfo {year}
  {2010})}\BibitemShut {NoStop}%
\bibitem [{\citenamefont {Neubauer}\ \emph {et~al.}(2009)\citenamefont
  {Neubauer}, \citenamefont {Pfleiderer}, \citenamefont {Binz}, \citenamefont
  {Rosch}, \citenamefont {Ritz}, \citenamefont {Niklowitz},\ and\ \citenamefont
  {B\"oni}}]{Neubauer2009}%
  \BibitemOpen
  \bibfield  {author} {\bibinfo {author} {\bibfnamefont {A.}~\bibnamefont
  {Neubauer}}, \bibinfo {author} {\bibfnamefont {C.}~\bibnamefont
  {Pfleiderer}}, \bibinfo {author} {\bibfnamefont {B.}~\bibnamefont {Binz}},
  \bibinfo {author} {\bibfnamefont {A.}~\bibnamefont {Rosch}}, \bibinfo
  {author} {\bibfnamefont {R.}~\bibnamefont {Ritz}}, \bibinfo {author}
  {\bibfnamefont {P.~G.}\ \bibnamefont {Niklowitz}}, \ and\ \bibinfo {author}
  {\bibfnamefont {P.}~\bibnamefont {B\"oni}},\ }\bibfield  {title} {\enquote
  {\bibinfo {title} {{Topological Hall Effect in the $A$ Phase of MnSi}},}\
  }\href {\doibase 10.1103/PhysRevLett.102.186602} {\bibfield  {journal}
  {\bibinfo  {journal} {Phys. Rev. Lett.}\ }\textbf {\bibinfo {volume} {102}},\
  \bibinfo {pages} {186602} (\bibinfo {year} {2009})}\BibitemShut {NoStop}%
\bibitem [{\citenamefont {Zang}\ \emph {et~al.}(2011)\citenamefont {Zang},
  \citenamefont {Mostovoy}, \citenamefont {Han},\ and\ \citenamefont
  {Nagaosa}}]{Zang2011}%
  \BibitemOpen
  \bibfield  {author} {\bibinfo {author} {\bibfnamefont {J.}~\bibnamefont
  {Zang}}, \bibinfo {author} {\bibfnamefont {M.}~\bibnamefont {Mostovoy}},
  \bibinfo {author} {\bibfnamefont {J.~H.}\ \bibnamefont {Han}}, \ and\
  \bibinfo {author} {\bibfnamefont {N.}~\bibnamefont {Nagaosa}},\ }\bibfield
  {title} {\enquote {\bibinfo {title} {Dynamics of skyrmion crystals in
  metallic thin films},}\ }\href {\doibase 10.1103/PhysRevLett.107.136804}
  {\bibfield  {journal} {\bibinfo  {journal} {Phys. Rev. Lett.}\ }\textbf
  {\bibinfo {volume} {107}},\ \bibinfo {pages} {136804} (\bibinfo {year}
  {2011})}\BibitemShut {NoStop}%
\bibitem [{\citenamefont {Lin}\ \emph {et~al.}(2013)\citenamefont {Lin},
  \citenamefont {Reichhardt}, \citenamefont {Batista},\ and\ \citenamefont
  {Saxena}}]{Saxena2013}%
  \BibitemOpen
  \bibfield  {author} {\bibinfo {author} {\bibfnamefont {S.-Z.}\ \bibnamefont
  {Lin}}, \bibinfo {author} {\bibfnamefont {C.}~\bibnamefont {Reichhardt}},
  \bibinfo {author} {\bibfnamefont {C.~D.}\ \bibnamefont {Batista}}, \ and\
  \bibinfo {author} {\bibfnamefont {A.}~\bibnamefont {Saxena}},\ }\bibfield
  {title} {\enquote {\bibinfo {title} {Particle model for skyrmions in metallic
  chiral magnets: Dynamics, pinning, and creep},}\ }\href {\doibase
  10.1103/PhysRevB.87.214419} {\bibfield  {journal} {\bibinfo  {journal} {Phys.
  Rev. B}\ }\textbf {\bibinfo {volume} {87}},\ \bibinfo {pages} {214419}
  (\bibinfo {year} {2013})}\BibitemShut {NoStop}%
\bibitem [{\citenamefont {Romming}\ \emph {et~al.}(2013)\citenamefont
  {Romming}, \citenamefont {Hanneken}, \citenamefont {Menzel}, \citenamefont
  {Bickel}, \citenamefont {Wolter}, \citenamefont {von Bergmann}, \citenamefont
  {Kubetzka},\ and\ \citenamefont {Wiesendanger}}]{Romming2013}%
  \BibitemOpen
  \bibfield  {author} {\bibinfo {author} {\bibfnamefont {N.}~\bibnamefont
  {Romming}}, \bibinfo {author} {\bibfnamefont {C.}~\bibnamefont {Hanneken}},
  \bibinfo {author} {\bibfnamefont {M.}~\bibnamefont {Menzel}}, \bibinfo
  {author} {\bibfnamefont {J.~E.}\ \bibnamefont {Bickel}}, \bibinfo {author}
  {\bibfnamefont {B.}~\bibnamefont {Wolter}}, \bibinfo {author} {\bibfnamefont
  {K.}~\bibnamefont {von Bergmann}}, \bibinfo {author} {\bibfnamefont
  {A.}~\bibnamefont {Kubetzka}}, \ and\ \bibinfo {author} {\bibfnamefont
  {R.}~\bibnamefont {Wiesendanger}},\ }\bibfield  {title} {\enquote {\bibinfo
  {title} {{Writing and Deleting Single Magnetic Skyrmions}},}\ }\href
  {\doibase 10.1126/science.1240573} {\bibfield  {journal} {\bibinfo  {journal}
  {Science}\ }\textbf {\bibinfo {volume} {341}},\ \bibinfo {pages} {636}
  (\bibinfo {year} {2013})}\BibitemShut {NoStop}%
\bibitem [{\citenamefont {von Bergmann}\ \emph {et~al.}(2014)\citenamefont {von
  Bergmann}, \citenamefont {Kubetzka}, \citenamefont {Pietzsch},\ and\
  \citenamefont {Wiesendanger}}]{Bergmann2014}%
  \BibitemOpen
  \bibfield  {author} {\bibinfo {author} {\bibfnamefont {K.}~\bibnamefont {von
  Bergmann}}, \bibinfo {author} {\bibfnamefont {A.}~\bibnamefont {Kubetzka}},
  \bibinfo {author} {\bibfnamefont {O.}~\bibnamefont {Pietzsch}}, \ and\
  \bibinfo {author} {\bibfnamefont {R.}~\bibnamefont {Wiesendanger}},\
  }\bibfield  {title} {\enquote {\bibinfo {title} {{Interface-induced chiral
  domain walls, spin spirals and skyrmions revealed by spin-polarized scanning
  tunneling microscopy}},}\ }\href {\doibase 10.1088/0953-8984/26/39/394002}
  {\bibfield  {journal} {\bibinfo  {journal} {J. Phys.: Condens. Matter}\
  }\textbf {\bibinfo {volume} {26}},\ \bibinfo {pages} {394002} (\bibinfo
  {year} {2014})}\BibitemShut {NoStop}%
\bibitem [{\citenamefont {Romming}\ \emph {et~al.}(2015)\citenamefont
  {Romming}, \citenamefont {Kubetzka}, \citenamefont {Hanneken}, \citenamefont
  {von Bergmann},\ and\ \citenamefont {Wiesendanger}}]{Romming2015}%
  \BibitemOpen
  \bibfield  {author} {\bibinfo {author} {\bibfnamefont {N.}~\bibnamefont
  {Romming}}, \bibinfo {author} {\bibfnamefont {A.}~\bibnamefont {Kubetzka}},
  \bibinfo {author} {\bibfnamefont {C.}~\bibnamefont {Hanneken}}, \bibinfo
  {author} {\bibfnamefont {K.}~\bibnamefont {von Bergmann}}, \ and\ \bibinfo
  {author} {\bibfnamefont {R.}~\bibnamefont {Wiesendanger}},\ }\bibfield
  {title} {\enquote {\bibinfo {title} {{Field-Dependent Size and Shape of
  Single Magnetic Skyrmions}},}\ }\href {\doibase
  10.1103/PhysRevLett.114.177203} {\bibfield  {journal} {\bibinfo  {journal}
  {Phys. Rev. Lett.}\ }\textbf {\bibinfo {volume} {114}},\ \bibinfo {pages}
  {177203} (\bibinfo {year} {2015})}\BibitemShut {NoStop}%
\bibitem [{\citenamefont {Fert}\ \emph {et~al.}(2013)\citenamefont {Fert},
  \citenamefont {Cros},\ and\ \citenamefont {Sampaio}}]{Fert2013}%
  \BibitemOpen
  \bibfield  {author} {\bibinfo {author} {\bibfnamefont {A.}~\bibnamefont
  {Fert}}, \bibinfo {author} {\bibfnamefont {V.}~\bibnamefont {Cros}}, \ and\
  \bibinfo {author} {\bibfnamefont {J.}~\bibnamefont {Sampaio}},\ }\bibfield
  {title} {\enquote {\bibinfo {title} {Skyrmions on the track},}\ }\href
  {\doibase 10.1038/nnano.2013.29} {\bibfield  {journal} {\bibinfo  {journal}
  {Nat. Nano}\ }\textbf {\bibinfo {volume} {8}},\ \bibinfo {pages} {152--156}
  (\bibinfo {year} {2013})}\BibitemShut {NoStop}%
\bibitem [{\citenamefont {Nagaosa}\ and\ \citenamefont
  {Tokura}(2013)}]{Nagaosa2013}%
  \BibitemOpen
  \bibfield  {author} {\bibinfo {author} {\bibfnamefont {N.}~\bibnamefont
  {Nagaosa}}\ and\ \bibinfo {author} {\bibfnamefont {Y.}~\bibnamefont
  {Tokura}},\ }\bibfield  {title} {\enquote {\bibinfo {title} {Topological
  properties and dynamics of magnetic skyrmions},}\ }\href
  {http://dx.doi.org/10.1038/nnano.2013.243} {\bibfield  {journal} {\bibinfo
  {journal} {Nat. Nanotechnol.}\ }\textbf {\bibinfo {volume} {8}},\ \bibinfo
  {pages} {899} (\bibinfo {year} {2013})}\BibitemShut {NoStop}%
\bibitem [{\citenamefont {Hurst}\ \emph {et~al.}(2015)\citenamefont {Hurst},
  \citenamefont {Efimkin}, \citenamefont {Zang},\ and\ \citenamefont
  {Galitski}}]{Hurst2015}%
  \BibitemOpen
  \bibfield  {author} {\bibinfo {author} {\bibfnamefont {H.~M.}\ \bibnamefont
  {Hurst}}, \bibinfo {author} {\bibfnamefont {D.~K.}\ \bibnamefont {Efimkin}},
  \bibinfo {author} {\bibfnamefont {J.}~\bibnamefont {Zang}}, \ and\ \bibinfo
  {author} {\bibfnamefont {V.}~\bibnamefont {Galitski}},\ }\bibfield  {title}
  {\enquote {\bibinfo {title} {{Charged skyrmions on the surface of a
  topological insulator}},}\ }\href {\doibase 10.1103/PhysRevB.91.060401}
  {\bibfield  {journal} {\bibinfo  {journal} {Phys. Rev. B}\ }\textbf {\bibinfo
  {volume} {91}},\ \bibinfo {pages} {060401} (\bibinfo {year}
  {2015})}\BibitemShut {NoStop}%
\bibitem [{\citenamefont {Alicea}(2012)}]{Alicea2012}%
  \BibitemOpen
  \bibfield  {author} {\bibinfo {author} {\bibfnamefont {J.}~\bibnamefont
  {Alicea}},\ }\bibfield  {title} {\enquote {\bibinfo {title} {{New directions
  in the pursuit of Majorana fermions in solid state systems.}}}\ }\href
  {\doibase 10.1088/0034-4885/75/7/076501} {\bibfield  {journal} {\bibinfo
  {journal} {Rep. Prog. Phys.}\ }\textbf {\bibinfo {volume} {75}},\ \bibinfo
  {pages} {076501} (\bibinfo {year} {2012})}\BibitemShut {NoStop}%
\bibitem [{\citenamefont {Beenakker}(2015)}]{Beenakker}%
  \BibitemOpen
  \bibfield  {author} {\bibinfo {author} {\bibfnamefont {C.~W.~J.}\
  \bibnamefont {Beenakker}},\ }\bibfield  {title} {\enquote {\bibinfo {title}
  {{Random-matrix theory of Majorana fermions and topological
  superconductors}},}\ }\href {\doibase 10.1103/RevModPhys.87.1037} {\bibfield
  {journal} {\bibinfo  {journal} {Rev. Mod. Phys.}\ }\textbf {\bibinfo {volume}
  {87}},\ \bibinfo {pages} {1037} (\bibinfo {year} {2015})}\BibitemShut
  {NoStop}%
\bibitem [{\citenamefont {Nayak}\ \emph {et~al.}(2008)\citenamefont {Nayak},
  \citenamefont {Simon}, \citenamefont {Stern}, \citenamefont {Freedman},\ and\
  \citenamefont {Das~Sarma}}]{Nayak2008}%
  \BibitemOpen
  \bibfield  {author} {\bibinfo {author} {\bibfnamefont {C.}~\bibnamefont
  {Nayak}}, \bibinfo {author} {\bibfnamefont {S.~H.}\ \bibnamefont {Simon}},
  \bibinfo {author} {\bibfnamefont {A.}~\bibnamefont {Stern}}, \bibinfo
  {author} {\bibfnamefont {M.}~\bibnamefont {Freedman}}, \ and\ \bibinfo
  {author} {\bibfnamefont {S.}~\bibnamefont {Das~Sarma}},\ }\bibfield  {title}
  {\enquote {\bibinfo {title} {Non-abelian anyons and topological quantum
  computation},}\ }\href {\doibase 10.1103/RevModPhys.80.1083} {\bibfield
  {journal} {\bibinfo  {journal} {Rev. Mod. Phys.}\ }\textbf {\bibinfo {volume}
  {80}},\ \bibinfo {pages} {1083--1159} (\bibinfo {year} {2008})}\BibitemShut
  {NoStop}%
\bibitem [{\citenamefont {Yu}(1965)}]{Yu}%
  \BibitemOpen
  \bibfield  {author} {\bibinfo {author} {\bibfnamefont {L.}~\bibnamefont
  {Yu}},\ }\bibfield  {title} {\enquote {\bibinfo {title} {{Bound state in
  superconductors with paramagnetic impurities}},}\ }\href {\doibase
  10.7498/aps.21.75} {\bibfield  {journal} {\bibinfo  {journal} {Acta Phys.
  Sin.}\ }\textbf {\bibinfo {volume} {21}},\ \bibinfo {pages} {75} (\bibinfo
  {year} {1965})}\BibitemShut {NoStop}%
\bibitem [{\citenamefont {Shiba}(1968)}]{Shiba}%
  \BibitemOpen
  \bibfield  {author} {\bibinfo {author} {\bibfnamefont {H.}~\bibnamefont
  {Shiba}},\ }\bibfield  {title} {\enquote {\bibinfo {title} {{Classical Spins
  in Superconductors}},}\ }\href {\doibase 10.1143/PTP.40.435} {\bibfield
  {journal} {\bibinfo  {journal} {Prog. Theor. Phys.}\ }\textbf {\bibinfo
  {volume} {40}},\ \bibinfo {pages} {435} (\bibinfo {year} {1968})}\BibitemShut
  {NoStop}%
\bibitem [{\citenamefont {Rusinov}(1969)}]{Rusinov}%
  \BibitemOpen
  \bibfield  {author} {\bibinfo {author} {\bibfnamefont {A.~I.}\ \bibnamefont
  {Rusinov}},\ }\bibfield  {title} {\enquote {\bibinfo {title}
  {Superconductivity near a paramagnetic impurity},}\ }\href
  {http://www.jetpletters.ac.ru/ps/1658/article_25295.shtml} {\bibfield
  {journal} {\bibinfo  {journal} {JETP Lett.}\ }\textbf {\bibinfo {volume}
  {9}},\ \bibinfo {pages} {85} (\bibinfo {year} {1969})}\BibitemShut {NoStop}%
\bibitem [{\citenamefont {Balatsky}\ \emph {et~al.}(2006)\citenamefont
  {Balatsky}, \citenamefont {Vekhter},\ and\ \citenamefont
  {Zhu}}]{Balatsky2006}%
  \BibitemOpen
  \bibfield  {author} {\bibinfo {author} {\bibfnamefont {A.~V.}\ \bibnamefont
  {Balatsky}}, \bibinfo {author} {\bibfnamefont {I.}~\bibnamefont {Vekhter}}, \
  and\ \bibinfo {author} {\bibfnamefont {J.-X.}\ \bibnamefont {Zhu}},\
  }\bibfield  {title} {\enquote {\bibinfo {title} {{Impurity-induced states in
  conventional and unconventional superconductors}},}\ }\href {\doibase
  10.1103/RevModPhys.78.373} {\bibfield  {journal} {\bibinfo  {journal} {Rev.
  Mod. Phys.}\ }\textbf {\bibinfo {volume} {78}},\ \bibinfo {pages} {373}
  (\bibinfo {year} {2006})}\BibitemShut {NoStop}%
\bibitem [{\citenamefont {Yao}\ \emph {et~al.}(2014)\citenamefont {Yao},
  \citenamefont {Glazman}, \citenamefont {Demler}, \citenamefont {Lukin},\ and\
  \citenamefont {Sau}}]{Yao2014}%
  \BibitemOpen
  \bibfield  {author} {\bibinfo {author} {\bibfnamefont {N.~Y.}\ \bibnamefont
  {Yao}}, \bibinfo {author} {\bibfnamefont {L.~I.}\ \bibnamefont {Glazman}},
  \bibinfo {author} {\bibfnamefont {E.~A.}\ \bibnamefont {Demler}}, \bibinfo
  {author} {\bibfnamefont {M.~D.}\ \bibnamefont {Lukin}}, \ and\ \bibinfo
  {author} {\bibfnamefont {J.~D.}\ \bibnamefont {Sau}},\ }\bibfield  {title}
  {\enquote {\bibinfo {title} {{Enhanced Antiferromagnetic Exchange between
  Magnetic Impurities in a Superconducting Host}},}\ }\href {\doibase
  10.1103/PhysRevLett.113.087202} {\bibfield  {journal} {\bibinfo  {journal}
  {Phys. Rev. Lett.}\ }\textbf {\bibinfo {volume} {113}},\ \bibinfo {pages}
  {087202} (\bibinfo {year} {2014})}\BibitemShut {NoStop}%
\bibitem [{\citenamefont {Garaud}\ \emph {et~al.}(2011)\citenamefont {Garaud},
  \citenamefont {Carlstr\"om},\ and\ \citenamefont {Babaev}}]{Garaud2011}%
  \BibitemOpen
  \bibfield  {author} {\bibinfo {author} {\bibfnamefont {J.}~\bibnamefont
  {Garaud}}, \bibinfo {author} {\bibfnamefont {J.}~\bibnamefont {Carlstr\"om}},
  \ and\ \bibinfo {author} {\bibfnamefont {E.}~\bibnamefont {Babaev}},\
  }\bibfield  {title} {\enquote {\bibinfo {title} {Topological solitons in
  three-band superconductors with broken time reversal symmetry},}\ }\href
  {\doibase 10.1103/PhysRevLett.107.197001} {\bibfield  {journal} {\bibinfo
  {journal} {Phys. Rev. Lett.}\ }\textbf {\bibinfo {volume} {107}},\ \bibinfo
  {pages} {197001} (\bibinfo {year} {2011})}\BibitemShut {NoStop}%
\bibitem [{\citenamefont {Nakosai}\ \emph {et~al.}(2013)\citenamefont
  {Nakosai}, \citenamefont {Tanaka},\ and\ \citenamefont
  {Nagaosa}}]{Nakosai2013}%
  \BibitemOpen
  \bibfield  {author} {\bibinfo {author} {\bibfnamefont {S.}~\bibnamefont
  {Nakosai}}, \bibinfo {author} {\bibfnamefont {Y.}~\bibnamefont {Tanaka}}, \
  and\ \bibinfo {author} {\bibfnamefont {N.}~\bibnamefont {Nagaosa}},\
  }\bibfield  {title} {\enquote {\bibinfo {title} {{Two-dimensional p-wave
  superconducting states with magnetic moments on a conventional s-wave
  superconductor}},}\ }\href {\doibase 10.1103/PhysRevB.88.180503} {\bibfield
  {journal} {\bibinfo  {journal} {Phys. Rev. B}\ }\textbf {\bibinfo {volume}
  {88}},\ \bibinfo {pages} {180503} (\bibinfo {year} {2013})}\BibitemShut
  {NoStop}%
\bibitem [{\citenamefont {Yokoyama}\ and\ \citenamefont
  {Linder}(2015)}]{Yokoyama2015}%
  \BibitemOpen
  \bibfield  {author} {\bibinfo {author} {\bibfnamefont {T.}~\bibnamefont
  {Yokoyama}}\ and\ \bibinfo {author} {\bibfnamefont {J.}~\bibnamefont
  {Linder}},\ }\bibfield  {title} {\enquote {\bibinfo {title} {Josephson effect
  through magnetic skyrmions},}\ }\href {\doibase 10.1103/PhysRevB.92.060503}
  {\bibfield  {journal} {\bibinfo  {journal} {Phys. Rev. B}\ }\textbf {\bibinfo
  {volume} {92}},\ \bibinfo {pages} {060503} (\bibinfo {year}
  {2015})}\BibitemShut {NoStop}%
\bibitem [{Note1()}]{Note1}%
  \BibitemOpen
  \bibinfo {note} {Note that for the case of a spin-singlet SC given by
  Eq.~(\ref {ham}), the Bloch and the Neel skyrmions are equivalent since they
  can be related by a continuous $\pi /2$-rotation around the $z$-axis in the
  spin space $U = \protect \qopname \relax o{exp}(-i\pi \sigma _z/4)$. In the
  presence of either the spin-triplet pairing or the spin-orbit interaction,
  the effects of the two types of skyrmions are different.}\BibitemShut {Stop}%
\bibitem [{Note2()}]{Note2}%
  \BibitemOpen
  \bibinfo {note} {Neglecting the orbital effect of the magnetic field is
  reasonable for ferromagnetic films of atomic thickness, for example produced
  in Hamburg~\cite {Romming2013,Bergmann2014,Romming2015}. On the other hand,
  the orbital effect of the magnetic field cannot be neglected for the 3D
  ferromagnets. In this case, the magnetic field produces vortices in a type-II
  superconductor. Note that interaction between a skyrmion and vortex was
  recently considered in Ref. \cite {Hals2016}.}\BibitemShut {Stop}%
\bibitem [{Note3()}]{Note3}%
  \BibitemOpen
  \bibinfo {note} {We expect that a different spatial dependency of the
  azimuthal angle (\ref {conf}) will not change the results
  significantly.}\BibitemShut {Stop}%
\bibitem [{\citenamefont {Braunecker}\ and\ \citenamefont
  {Simon}(2013)}]{Braunecker2013}%
  \BibitemOpen
  \bibfield  {author} {\bibinfo {author} {\bibfnamefont {B.}~\bibnamefont
  {Braunecker}}\ and\ \bibinfo {author} {\bibfnamefont {P.}~\bibnamefont
  {Simon}},\ }\bibfield  {title} {\enquote {\bibinfo {title} {{Interplay
  between Classical Magnetic Moments and Superconductivity in Quantum
  One-Dimensional Conductors: Toward a Self-Sustained Topological Majorana
  Phase}},}\ }\href {\doibase 10.1103/PhysRevLett.111.147202} {\bibfield
  {journal} {\bibinfo  {journal} {Phys. Rev. Lett.}\ }\textbf {\bibinfo
  {volume} {111}},\ \bibinfo {pages} {147202} (\bibinfo {year}
  {2013})}\BibitemShut {NoStop}%
\bibitem [{\citenamefont {Klinovaja}\ \emph {et~al.}(2013)\citenamefont
  {Klinovaja}, \citenamefont {Stano}, \citenamefont {Yazdani},\ and\
  \citenamefont {Loss}}]{Klinovaja2013}%
  \BibitemOpen
  \bibfield  {author} {\bibinfo {author} {\bibfnamefont {J.}~\bibnamefont
  {Klinovaja}}, \bibinfo {author} {\bibfnamefont {P.}~\bibnamefont {Stano}},
  \bibinfo {author} {\bibfnamefont {A.}~\bibnamefont {Yazdani}}, \ and\
  \bibinfo {author} {\bibfnamefont {D.}~\bibnamefont {Loss}},\ }\bibfield
  {title} {\enquote {\bibinfo {title} {{Topological Superconductivity and
  Majorana Fermions in RKKY Systems}},}\ }\href {\doibase
  10.1103/PhysRevLett.111.186805} {\bibfield  {journal} {\bibinfo  {journal}
  {Phys. Rev. Lett.}\ }\textbf {\bibinfo {volume} {111}},\ \bibinfo {pages}
  {186805} (\bibinfo {year} {2013})}\BibitemShut {NoStop}%
\bibitem [{\citenamefont {Vazifeh}\ and\ \citenamefont
  {Franz}(2013)}]{Vazifeh2013}%
  \BibitemOpen
  \bibfield  {author} {\bibinfo {author} {\bibfnamefont {M.~M.}\ \bibnamefont
  {Vazifeh}}\ and\ \bibinfo {author} {\bibfnamefont {M.}~\bibnamefont
  {Franz}},\ }\bibfield  {title} {\enquote {\bibinfo {title} {{Self-Organized
  Topological State with Majorana Fermions}},}\ }\href {\doibase
  10.1103/PhysRevLett.111.206802} {\bibfield  {journal} {\bibinfo  {journal}
  {Phys. Rev. Lett.}\ }\textbf {\bibinfo {volume} {111}},\ \bibinfo {pages}
  {206802} (\bibinfo {year} {2013})}\BibitemShut {NoStop}%
\bibitem [{Note4()}]{Note4}%
  \BibitemOpen
  \bibinfo {note} {Indeed, after the submission of our manuscript,
  Reference~\cite {Yang2016} found a zero energy Majorana bound state at the
  core of the skyrmion of a higher winding number.}\BibitemShut {Stop}%
\bibitem [{Note5()}]{Note5}%
  \BibitemOpen
  \bibinfo {note} {The domain of applicability can also be extended to
  $p_F^{-1}<R\ll \xi _{sc}$ with some modification of the theory.}\BibitemShut
  {Stop}%
\bibitem [{Note6()}]{Note6}%
  \BibitemOpen
  \bibinfo {note} {Although Eq.~(\ref {energy}) suggests that the SBS states
  may go inside the actual gap for large enough $S_0$, i.e. $|E^\pm _{\protect
  \rm SBS}|<\Delta -S$, the multipole approximation of a skyrmion (\ref {vr})
  is no-longer valid in this regime and, thus, does not give a reliable
  estimate of the SBS energy. In practice, by performing a numerical modeling,
  we never observe the SBS peaks inside the actual spectral gap, i.e. in the
  window of energies $|E^\pm _{\protect \rm SBS}|<\Delta -S$.}\BibitemShut
  {Stop}%
\bibitem [{Note7()}]{Note7}%
  \BibitemOpen
  \bibinfo {note} {The function $I_\psi ^{-1}$ is commonly referred to as the
  inverse participation ratio in the literature on localization
  phenomena.}\BibitemShut {Stop}%
\bibitem [{\citenamefont {M\'{e}nard}\ \emph {et~al.}(2015)\citenamefont
  {M\'{e}nard}, \citenamefont {Guissart}, \citenamefont {Brun}, \citenamefont
  {Pons}, \citenamefont {Stolyarov}, \citenamefont {Debontridder},
  \citenamefont {Leclerc}, \citenamefont {Janod}, \citenamefont {Cario},
  \citenamefont {Roditchev}, \citenamefont {Simon},\ and\ \citenamefont
  {Cren}}]{Menard2015}%
  \BibitemOpen
  \bibfield  {author} {\bibinfo {author} {\bibfnamefont {G.~C.}\ \bibnamefont
  {M\'{e}nard}}, \bibinfo {author} {\bibfnamefont {S.}~\bibnamefont
  {Guissart}}, \bibinfo {author} {\bibfnamefont {C.}~\bibnamefont {Brun}},
  \bibinfo {author} {\bibfnamefont {S.}~\bibnamefont {Pons}}, \bibinfo {author}
  {\bibfnamefont {V.~S.}\ \bibnamefont {Stolyarov}}, \bibinfo {author}
  {\bibfnamefont {F.}~\bibnamefont {Debontridder}}, \bibinfo {author}
  {\bibfnamefont {M.~V.}\ \bibnamefont {Leclerc}}, \bibinfo {author}
  {\bibfnamefont {E.}~\bibnamefont {Janod}}, \bibinfo {author} {\bibfnamefont
  {L.}~\bibnamefont {Cario}}, \bibinfo {author} {\bibfnamefont
  {D.}~\bibnamefont {Roditchev}}, \bibinfo {author} {\bibfnamefont
  {P.}~\bibnamefont {Simon}}, \ and\ \bibinfo {author} {\bibfnamefont
  {T.}~\bibnamefont {Cren}},\ }\bibfield  {title} {\enquote {\bibinfo {title}
  {{Coherent long-range magnetic bound states in a superconductor}},}\ }\href
  {http://dx.doi.org/10.1038/nphys3508} {\bibfield  {journal} {\bibinfo
  {journal} {Nat. Phys.}\ }\textbf {\bibinfo {volume} {11}},\ \bibinfo {pages}
  {1013} (\bibinfo {year} {2015})}\BibitemShut {NoStop}%
\bibitem [{\citenamefont {Yang}\ \emph {et~al.}(2016)\citenamefont {Yang},
  \citenamefont {Stano}, \citenamefont {Klinovaja},\ and\ \citenamefont
  {Loss}}]{Yang2016}%
  \BibitemOpen
  \bibfield  {author} {\bibinfo {author} {\bibfnamefont {G.}~\bibnamefont
  {Yang}}, \bibinfo {author} {\bibfnamefont {P.}~\bibnamefont {Stano}},
  \bibinfo {author} {\bibfnamefont {J.}~\bibnamefont {Klinovaja}}, \ and\
  \bibinfo {author} {\bibfnamefont {D.}~\bibnamefont {Loss}},\ }\bibfield
  {title} {\enquote {\bibinfo {title} {{Majorana bound states in magnetic
  skyrmions}},}\ }\href {\doibase 10.1103/PhysRevB.93.224505} {\bibfield
  {journal} {\bibinfo  {journal} {Phys. Rev. B}\ }\textbf {\bibinfo {volume}
  {93}},\ \bibinfo {pages} {224505} (\bibinfo {year} {2016})}\BibitemShut
  {NoStop}%
\bibitem [{\citenamefont {Hals}\ \emph {et~al.}(2016)\citenamefont {Hals},
  \citenamefont {Schecter},\ and\ \citenamefont {Rudner}}]{Hals2016}%
  \BibitemOpen
  \bibfield  {author} {\bibinfo {author} {\bibfnamefont {K.~M.~D.}\
  \bibnamefont {Hals}}, \bibinfo {author} {\bibfnamefont {M.}~\bibnamefont
  {Schecter}}, \ and\ \bibinfo {author} {\bibfnamefont {M.~S.}\ \bibnamefont
  {Rudner}},\ }\bibfield  {title} {\enquote {\bibinfo {title} {Composite
  topological excitations in ferromagnet-superconductor heterostructures},}\
  }\href {\doibase 10.1103/PhysRevLett.117.017001} {\bibfield  {journal}
  {\bibinfo  {journal} {Phys. Rev. Lett.}\ }\textbf {\bibinfo {volume} {117}},\
  \bibinfo {pages} {017001} (\bibinfo {year} {2016})}\BibitemShut {NoStop}%
\bibitem [{\citenamefont {Shytov}\ \emph {et~al.}(2009)\citenamefont {Shytov},
  \citenamefont {Abanin},\ and\ \citenamefont {Levitov}}]{Shytov2009}%
  \BibitemOpen
  \bibfield  {author} {\bibinfo {author} {\bibfnamefont {A.~V.}\ \bibnamefont
  {Shytov}}, \bibinfo {author} {\bibfnamefont {D.~A.}\ \bibnamefont {Abanin}},
  \ and\ \bibinfo {author} {\bibfnamefont {L.~S.}\ \bibnamefont {Levitov}},\
  }\bibfield  {title} {\enquote {\bibinfo {title} {Long-range interaction
  between adatoms in graphene},}\ }\href {\doibase
  10.1103/PhysRevLett.103.016806} {\bibfield  {journal} {\bibinfo  {journal}
  {Phys. Rev. Lett.}\ }\textbf {\bibinfo {volume} {103}},\ \bibinfo {pages}
  {016806} (\bibinfo {year} {2009})}\BibitemShut {NoStop}%
\bibitem [{\citenamefont {Abanin}\ and\ \citenamefont
  {Pesin}(2011)}]{Abanin2011}%
  \BibitemOpen
  \bibfield  {author} {\bibinfo {author} {\bibfnamefont {D.~A.}\ \bibnamefont
  {Abanin}}\ and\ \bibinfo {author} {\bibfnamefont {D.~A.}\ \bibnamefont
  {Pesin}},\ }\bibfield  {title} {\enquote {\bibinfo {title} {Ordering of
  magnetic impurities and tunable electronic properties of topological
  insulators},}\ }\href {\doibase 10.1103/PhysRevLett.106.136802} {\bibfield
  {journal} {\bibinfo  {journal} {Phys. Rev. Lett.}\ }\textbf {\bibinfo
  {volume} {106}},\ \bibinfo {pages} {136802} (\bibinfo {year}
  {2011})}\BibitemShut {NoStop}%
\end{thebibliography}%

\appendix

\section{Derivation of the T-matrix} \label{sec:appendixTMatrix}

In this section, we provide details on the derivation of the T-matrix~(\ref{tm1}) for the model given by Eqs.~(\ref{ham}) and (\ref{vr1}). In the momentum space, the local terms defined via the delta functions in Eq.~(\ref{vr1}) generate the following perturbation
\begin{equation}
	V(\bm p) = -S_0\,\sigma_z +  i \,S_1 \, \bm \sigma\cdot \bm  p.
	\label{vp}
\end{equation}
Using Eq.~(\ref{vp}) and the bare Green's function~(\ref{grf}), we write the Lippmann-Schwinger integral equation for the T-matrix
\begin{align}
	T\left(\bm p^{\rm out},\bm p^{\rm in}\right) &= V \left(\bm p^{\rm out}-\bm p^{\rm in}\right) + \\
	&\int \frac{d^2 p'}{\left( 2\pi \right)^2}\, V\left(\bm p^{\rm out}-\bm p'\right) g(\omega,\bm p')  T\left(\bm p',\bm p^{\rm in}\right).
	\label{integEq}
\end{align}
We focus on the scatering close to the Fermi surface, so we use $\bm p^{\rm out} = p_F\, \bm n^{\rm out}$ and $\bm p^{\rm in} = p_F \,\bm n^{\rm in}$, where $\bm n^{\rm out}$ and $\bm n^{\rm in}$ the in-plane unit vectors. Then, we choose the following  ansatz for the T-matrix 
\begin{align}
	T\left(\bm n^{\rm out},\bm n^{\rm in}\right) &= T^0 + T^1_i n^{\rm out}_i + {T^1_j}^\dagger n^{\rm in}_j + T^2_{ij}\, n^{\rm out}_i n^{\rm in}_j, \label{ansatz}
\end{align}
where  $T,T^1_i,T^2_{ij}$ are the 4-by-4 matrices in the $\sigma\otimes\tau$ space, which give the expansion of the T-matrix in vectors $n^{\rm out}_i$ and $n^{\rm in}_j$. We substitute the ansatz~(\ref{ansatz}) in Eq.~(\ref{integEq}) and rewrite the integral equation as 
\begin{align}
	& T\left(\bm n^{\rm out},\bm n^{\rm in}\right) =  -S_0\,\sigma_z +  i \,S_1p_F \, \sigma_i \left(  n^{\rm out}_i-n^{\rm in}_i \right) +  \nonumber\\
	&  \int d\bm n' \left[ -S_0\,\sigma_z +  i \,S_1p_F \, \sigma_i \left(  n^{\rm out}_i-n'_i \right) \right] g_0(\omega) \times \nonumber \\
	& \quad\left[ T^0 + T^1_j n'_j + {T^1_j}^\dagger n^{\rm in}_j + T^2_{kj}\, n'_k n^{\rm in}_j \right],
	\label{integEq1}
\end{align}
where after an integration in the radial variable $p'$ the Green's function in the momentum space $g(\omega,\bm p')$ transformed into an on-site matrix of the Green's functions $g_0(\omega)$. Next, we take an integral over the angular variable $\bm n'$, i.e. $\int d\bm n'\, n'_i\, = 0$ and $\int d\bm n'\, n'_in'_j\, = \delta_{ij}/2$, and obtain a closed set of equations for the unknown matrices
\begin{align}
	T^0 &= -S_0\sigma_z-S_0\sigma_zg_0\,T^0-\frac12 iS_1p_F\,\sigma_ig_0\,T^1_i, \label{eq0} \\
	T^1_i &= iS_1p_F\,\sigma_i\left[1+g_0(\omega)\,T^0\right], \label{eq1} \\
	T^2_{ij} &= iS_1p_F\,\sigma_ig_0(\omega)\,{T^1_j}^\dagger. \label{eq2}
\end{align}
Solution of the Eqs.~(\ref{eq0})-(\ref{eq2}) gives
\begin{align}
	& T^0 = \left[ -S_0\sigma_z+S_1^2p_F^2\bar g_0(\omega) \right] D \nonumber,\\ 
	& T^1_i = iS_0p_F\,\sigma_i\,D, \nonumber\\ 
	& T^2_{ij} = S_1^2p_F^2\,\sigma_ig_0D\sigma_j,  \label{TM} \\
	& {\rm where}\,D = \left[ 1+S_0\sigma_zg_{0}-S^2_1p_F^2\,\bar g_{0}(\omega)\, g_{0}(\omega) \right]^{-1}. \nonumber
\end{align}
Note that the relative order of the matrices in Eq.~(\ref{TM}) is important because the spin Pauli matrices do not commute. For brevity, $\bar g_0(\omega) = \frac{1}{2}\sum_{j=x,y} \sigma_j g_0(\omega) \sigma_j $ denotes the Green's function obtained from $g_{0}$ by replacing $\sigma_z \rightarrow - \sigma_z$. So, in the presence of the skyrmion, the Green's function becomes
\begin{align}
	G(\omega,\bm p^1,\bm p^2) =& g(\omega,\bm p^1)\,(2\pi)^2\delta(\bm p^1-\bm p^2) + \nonumber \\
& \quad	g(\omega,\bm p^1) T(\bm p^1,\bm p^2) g(\omega,\bm p^2), 	\label{G}
\end{align}
using which the spin-polarized local density of states (SP-LDOS) can be expressed
\begin{align}
	&\rho_s(\omega,\bm r) =  \\
	-&\frac{1}{\pi}\,{\rm Im\,Tr} \left[ \frac{1+\tau_z}{2}\,\frac{1+\sigma_s}{2} \int \frac{d^2p^1\,d^2p^2}{\left( 2\pi \right)^4} e^{i\left( \bm p^1-\bm p^2 \right)\bm r} G(\omega,\bm p^1,\bm p^2)\right] \label{rhor}
\end{align}
where $s=x,y,z$ denotes the spin quantization axis. At the skyrmion core, i.e. at $\bm r=0$, only the $T^0$ part of the T-matrix contributes to local density of states 
\begin{align}
	& \rho_s(\omega,0) =  -\frac{1}{\pi}\,{\rm Im \,Tr} \left\{  \frac{1+\tau_z}{2}\,\frac{1+\sigma_s}{2}  \left[g_0(\omega)+\right.\right. \label{spldos}\\
	& \left.\left. g_0(\omega)  \frac{-S_0\sigma_z+S^2_1p_F^2\bar g_{0}(\omega)}{1+S_0\sigma_zg_{0}(\omega)-S^2_1p_F^2\,\bar g_{0}(\omega)\, g_{0}(\omega)} g_0(\omega)  \right]\right\}\, \nonumber 
\end{align}
whereas  $T^1$ and $T^2$ drop out. Equation~(\ref{spldos}) gives the expression in Eq.~(12).

\section{Interaction between skyrmions} \label{sec:appendixInteraction}

\begin{figure} \centering
	\includegraphics[width=0.7\linewidth]{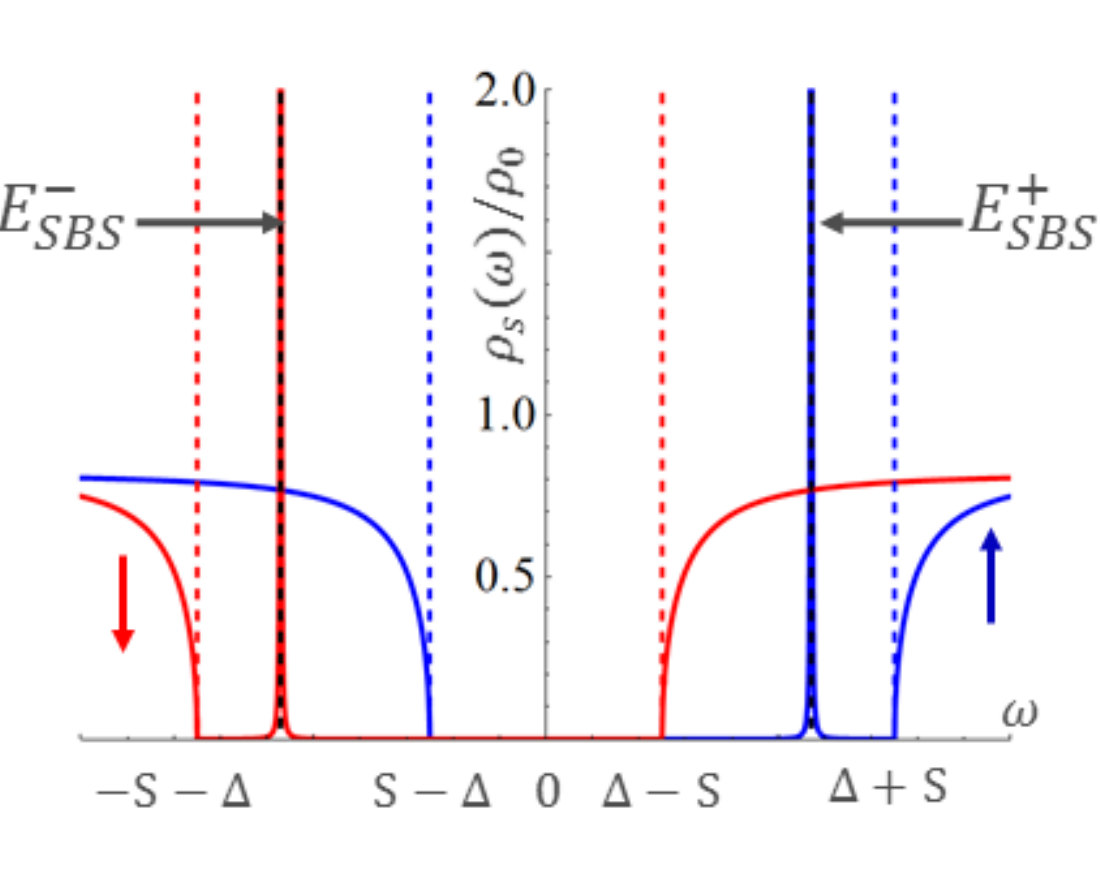}	
	\caption{(color online) Spin-polarized local density of states in the absence of the in-plane spins, i.e. at $S_1 = 0$. The localized skyrmion bound states do not couple to the delocalized states of the opposite spin-polarization and are described by the sharp poles given by Eq.~(\ref{tmat}). Blue and red colors encode the up and down spin polarizaion.} \label{fig:SPLDOS}
\end{figure}

In order to estimate the superconductivity-induced interaction between skyrmions we use the formalism of TGTG formula usually discussed in the context of the Casimir interaction. It was also recently used in the condensed matter context to describe the interaction between impurities in graphene~\cite{Shytov2009} and topological insulators~\cite{Abanin2011} mediated by electrons. At zero temperature $T=0$, the free energy interaction between the skyrmions can be expressed as  
\begin{equation}
	U_{\rm int}(\bm r) = \frac{1}{\pi}\int_{-\infty}^0 d\omega\,\, {\rm Im\,Tr\, Log\,} \left[ 1- g_{\bm r}(\omega) T_1(\omega) g_{-\bm r}(\omega) T_2(\omega) \right],
	\label{interaction}
\end{equation}
where the integral is taken over all negative energies, i.e. filled states, in the Bogolyubov-de Gennes formulation. In Eq.~(\ref{interaction}), $T_1$ and $T_2$ are the T-matrices corresponding to individual skyrmions, and the Green's function $g_{\bm r}$ is calculated in the real space for large $r\gtrsim\xi_{\rm sc}\gg p_F^{-1}$ 
\begin{widetext}
\begin{equation}
	g_{\bm r}(\omega) = - \sqrt{\frac{2\pi}{p_Fr}}\sum_{\lambda = \pm 1} \frac{1+\lambda\sigma_z}{2}\,\left[ \tau_z\cos\left( p_Fr+\frac{\pi}{4} \right) +\frac{\omega-\lambda S+\Delta\tau_x}{\sqrt{\Delta^2-\left( \omega-\lambda S \right)^2}}\sin\left( p_Fr+\frac{\pi}{4} \right) \right]\, \rho_0\, e^{-r\sqrt{\Delta^2-\left( \omega-\lambda S \right)^2}/v_F}.
	\label{grfunc}
\end{equation}
\end{widetext}
Here, the projector $ \frac{1+\lambda\sigma_z}{2}$ selects the spin-up and down sectors of the Hamiltonian, which are shifted in energy due to the constant FM field $-S\,\hat{\bm z}$ as discussed in the paper. Note a square root term in the exponent of Eq.~(\ref{grfunc}). For $|\omega-\lambda S|<\Delta$, the exponent produces an exponential decay at the scale of $r \sim v_F/\sqrt{\Delta^2-\left( \omega-\lambda S \right)^2}$. In contrast, the square root becomes purely imaginary $-i\,{\rm sgn}(\omega-\lambda S) \sqrt{\left( \omega-\lambda S \right)^2-\Delta^2}$ for $|\omega-\lambda S|>\Delta$, and so the exponential term gives periodic oscillations rather than exponential decay. This observation motivates the explanation of the long-range coupling: the skyrmion bound states couple to the delocalized states of the opposite spin polarization, for which the square root is imaginary. Then the Green's function, which has a long-range power-law behavior, can propagate between the skyrmions at large distances $r>\xi$ and, thus, couple their bound states and generate an effective interaction between skyrmions. 

The T-matrix given in Eq.~(\ref{TM}) has a complicated form. So, for simplicity, let us demonstrate the long-range interaction between skyrmions perturbatively in $S_1$. First, at $S_1 = 0$, i.e. where the in-plane scattering is neglected, the T-matrix~(\ref{TM}) reduces to a simpler Eq.~(8) of the main text. The corresponding SP LDOS is shown in Fig.~(\ref{fig:SPLDOS}). The skyrmion bound states are represented by the sharp peaks in the density of states which lie in the energy windows between the Zeeman split coherence peaks, i.e. $\Delta+S>|E^\pm_{\rm SBS}|>\Delta-S$. In this approximation, the scattering by the in-plane spins is absent, and, therefore, the localized states do not couple to the delocalized states of the opposite spin-polarization. Now, we consider the higher-order order terms of the T-matrix expansion in $S_1$. We look for the terms that would couple the skyrmion states to the delocalized states of the opposite spin polarization. In the second-order in $S_1$, there is one such term generated by the contribution $T^2_{ij}$ in Eq.~(\ref{TM}). So, in the vicinity of the energy close to the bound states energies, the relevant part of the T-matrix can be written as 
\begin{equation}
	\sigma_i n_i^{\rm out}\left[S_1^2\,\sum_{\lambda=\pm 1} \frac{1+\lambda\tau_x}{2}\frac{1+\lambda\sigma_z}{2} \,\,\frac{\alpha}{\omega-E^{\lambda}_{\rm SBS}}\right] \sigma_j n_j^{\rm in}
	\label{tmat}
\end{equation}
where the terms $\frac{1+\lambda\tau_x}{2}$ and $\frac{1+\lambda\sigma_z}{2}$ are the projectors in the Nambu and spin space, whereas constant $\alpha$ gives a strength of the bound state poles. Observe that Eq.~(\ref{tmat})  is dressed with the in-plane Pauli matrices $\sigma_i$ on both sides of the expression. The in-plane Pauli matrices $\sigma_i=(\sigma_x,\sigma_y)$ flip the spin $\sigma_z$ and, thus, couple the bound states poles to the background delocalized states. Then, we substitute Eqs.~(\ref{grfunc}) and (\ref{tmat}) in Eq.~(\ref{interaction}). The integral in Eq.~(\ref{interaction}) is dominated by the poles in the T-matrix, so we approximate the integrand as 
\begin{widetext}
\begin{align}
	{\rm Im\,Tr\, Log\,} \left[ 1-   \beta\frac{S_1^4}{p_Fr} \sum_{\lambda = \pm 1} \frac{1+\lambda\sigma_z}{2} \frac{1}{\left(\omega-E^{-\lambda}_{\rm SBS}\right)^2} e^{-i\lambda r\sqrt{\left( E^{-\lambda}_{\rm SBS}-\lambda S \right)^2-\Delta^2}/v_F} \right],
	\label{tgtg}
\end{align}
\end{widetext}
where $\beta$ is a constant absorbing other parameters. Observe, that the argument in the exponent is imaginary. Since the integral in Eq.~(\ref{interaction}) runs over negative energies, $\lambda=1$ dominates the integral and we focus only at the vicinity of $\omega$ around $E^{-}_{\rm SBS}$. So, after shifting the integration variable $\omega-E^{-}_{\rm SBS}\rightarrow\omega$, and reexpressing the imaginary part of the logarithm, we rewrite the integral as  
\begin{align}
	U_{\rm int}(r) = \frac{1}{\pi}\int_{-\infty}^\infty d\omega\, {\rm atan}\left[\frac{\sin \kappa r}{\frac{p_Fr}{\beta S_1^4}\omega^2-\cos \kappa r}  \right] = S_1^2\sqrt{\frac{\beta}{p_Fr}} I(\kappa r), 	\label{ur},
\end{align}
where $\kappa = \sqrt{\left( E^{-}_{\rm SBS}-S \right)^2-\Delta^2}/v_F$, and $I(r)=\frac{1}{\pi}\int_{-\infty}^\infty dx\, {\rm atan}\left[\frac{\sin \kappa r}{x^2-\cos \kappa r}  \right]$ is a periodic function of $\kappa r$: $I(\kappa r)= 2 \cos\left( \frac{\kappa r}{2} \right)$ for $4\pi n+2\pi>\kappa r>4\pi n$, and $I(\kappa r)= - 2 \cos\left( \frac{\kappa r}{2} \right)$ for $4\pi n+4\pi>\kappa r>4\pi n + 2\pi$. So, we find that interaction between skyrmions~(\ref{ur}) decays as $1/\sqrt{r}$ and oscillates at a scale of $1/\kappa$. In colloquial terms, the oscillating long-range wavefunctions corresponding to distinct skyrmions determine the effective interaction between skyrmions. Note that, we have calculated the contribution to the energy only due to the subgap states, and neglected the supragap states. The full analysis using Eqs.~(\ref{TM}) and (\ref{interaction}) will be given in a subsequent work.   

\end{document}